# Artificial intelligence for abnormality detection in high volume neuroimaging: a systematic review and meta-analysis

Siddharth Agarwal[1], David Wood[1], Mariusz Grzeda[1], Chandhini Suresh[2], Munaib Din[1], James Cole[3], Marc Modat[1], Thomas C Booth[1,4*]

1. School of Biomedical Engineering & Imaging Sciences, King's College London, Rayne Institute, 4th Floor, Lambeth Wing, London SE1 7EH, UK
2. Leicester Medical School, University of Leicester, Leicester LE1 7RH, UK
3. Centre for Medical Image Computing, Department of Computer Science, University College London, London WC1V 6LJ, UK
4. Department of Neuroradiology, Ruskin Wing, King's College Hospital NHS Foundation Trust, London SE5 9RS, UK

Siddharth Agarwal MBBS MRes FHEA, PhD Candidate, King's College London

David Wood PhD, PhD, Post-doctoral researcher, King's College London

Mariusz Grzeda PhD, Statistician, King's College London

Chandhini Suresh MPh, Medical Student, University of Leicester

Munaib Din MBBS, Academic Foundation Doctor, King's College London

James Cole PhD, Associate Professor in Neuroimage Analysis, University College London

Marc Modat PhD, Senior Lecturer, King's College London

Thomas C Booth EDiNR DMCC MA MBBS MRCP FRCR PhD, Reader in Neuroimaging, Kings College London, Consultant Diagnostic and Interventional Neuroradiologist, King's College Hospital

*Correspondence to: Thomas C Booth, thomas.booth@kcl.ac.uk, +44 (0) 203 299 4828

## ABSTRACT

Purpose: Most studies evaluating artificial intelligence (AI) models that detect abnormalities in neuroimaging are either tested on unrepresentative patient cohorts or are insufficiently well-validated, leading to poor generalisability to real-world tasks. The aim was to determine the diagnostic test accuracy and summarise the evidence supporting the use of AI models performing first-line, high-volume neuroimaging tasks.

Methods: Medline, Embase, Cochrane library and Web of Science were searched until September 2021 for studies that temporally or externally validated AI capable of detecting abnormalities in first-line CT or MR neuroimaging. A bivariate random-effects model was used for meta-analysis where appropriate. PROSPERO: CRD42021269563.

Results: Only 16 studies were eligible for inclusion. Included studies were not compromised by unrepresentative datasets or inadequate validation methodology. Direct comparison with radiologists was available in 4/16 studies. 15/16 had a high risk of bias. Meta-analysis was only suitable for intracranial haemorrhage detection in CT imaging (10/16 studies), where AI systems had a pooled sensitivity and specificity 0·90 (95% CI 0·85 - 0·94) and 0·90 (95% CI 0·83 - 0·95) respectively. Other AI studies using CT and MRI detected target conditions other than haemorrhage (2/16), or multiple target conditions (4/16). Only 3/16 studies implemented AI in clinical pathways, either for pre-read triage or as post-read discrepancy identifiers.

Conclusion: The paucity of eligible studies reflects that most abnormality detection AI studies were not adequately validated in representative clinical cohorts. The few studies describing how abnormality detection AI could impact patients and clinicians did not explore the full ramifications of clinical implementation.

## Competing interests
All authors have completed the ICMJE uniform disclosure form at www.icmje.org/coi_disclosure.pdf and declare: no support from any organisation for the submitted work; no financial relationships with any organisations that might have an interest in the submitted work in the previous three years; no other relationships or activities that could appear to have influenced the submitted work.

## Acknowledgements
SA is supported by an Engineering and Physical Sciences Research Council (EPSRC) funded PhD studentship (EP/R513064/1). This research was also supported by the Wellcome/EPSRC Centre for Medical Engineering (WT 203148/Z/16/Z) (TB, MG, MM) which includes open access fees, The Royal College of Radiologists (TB) and King's College Hospital Research and Innovation (TB).

## Author contributions

SA, DW, JC, MM and TCB conceived the study. SA, DW, and CS executed the search and extracted data. MG conceived and performed the meta-analysis and meta-regression. SA and TCB wrote the paper. All authors contributed to revisions of the manuscript and approved the final version. TCB is the study guarantor. The corresponding author attests that all listed authors meet authorship criteria and that no others meeting the criteria have been omitted.

## Data availability
Raw data are available on request from the corresponding author. The lead author and manuscript's guarantor (TCB) affirm that the manuscript is an honest, accurate, and transparent account of the study being reported; that no important aspects of the study have been omitted; and that any discrepancies from the study as planned (and, if relevant, registered) have been explained.

# INTRODUCTION

In the developed world, first-line imaging is performed in almost all hospitals, and refers to imaging performed at onset, for example, a CT head for an unconscious patient in the emergency department, or a head MRI for a patient with headache. First-line imaging is a high-volume task and a range of pathologies can be encountered. We distinguish this from second-line imaging where detailed biomarkers are extracted, based on prior clinical and first-line imaging information. Typically, second-line imaging is only performed in specialist hospitals where examples include large vessel occlusion imaging for stratifying stroke patients for thrombectomy treatment, or perfusion imaging for characterising brain tumours [1]. In comparison to first-line imaging, second-line imaging is a low volume task.

Radiology workloads for first-line imaging have soared in the last decade due to changing demographics, increased screening for early diagnosis initiatives, and updated clinical pathway guidelines requiring imaging. In the years leading up to the coronavirus disease 2019 (COVID-19) pandemic, the number of brain MRI scans performed in, for example, the UK increased on average by 7.8% annually, and the demand for both CT and MRI reporting outpaced the growth in the radiology workforce [2,3]. Reporting backlogs are problems of national importance in the UK, and analogous scenarios are seen in healthcare systems globally. Diagnostic delays cause poorer short and long-term clinical outcomes, with the late detection of illness inflating healthcare costs [4].

The automated detection of abnormalities in a scan using artificial intelligence (AI) has the potential to improve radiologist efficiency. AI can be used to reorder radiology worklists by flagging abnormal scans, as a reader aid or even as a second reader to identify missed pathology. However, the considerable interest in introducing AI into clinical environments to improve productivity in the high volume first-line imaging tasks, may be clouded by two main challenges in most published studies. Firstly, many abnormality detection AI studies report the diagnostic accuracy using non-representative clinical datasets (e.g. intracranial haemorrhage alone versus healthy controls without any other pathology) [5], including commercially available AI solutions [6]. Indeed, few studies validate their findings on datasets that are representative of the scans seen in routine clinical practice which contain a wide variety of pathologies. A second concern is that many studies do not demonstrate the generalisability of AI models due to inadequate validation methodology [7,8]. By validating abnormality detection AI on a hold-out subset from the same patient dataset, known as internal validation, it is unclear whether reported AI performance would translate to different patient populations scanned at different institutions. A recent systematic review analysing the enormous number of recent studies where AI was used for the detection of COVID-19 using chest imaging, found that all 62 included studies had no potential clinical use due to methodological biases such as the use of unrepresentative datasets and insufficient validation [9].

The aim of this systematic review was to determine the diagnostic performance and summarise the evidence supporting the use of those AI models carrying out first-line neuroimaging tasks. Critically, we ensured that our analyses were only focused on those studies that were not compromised by unrepresentative datasets or inadequate validation methodology. Therefore, we analysed those AI models that might conceivably be ready for use in the clinic. The primary objective was to determine the diagnostic accuracy of these AI models. A secondary objective was to determine the impact of AI on downstream clinical outcomes in those studies where this had been investigated.

# METHODS

This systematic review was conducted in accordance with the Preferred Reporting Items for Systematic Reviews and Meta-Analyses of Diagnostic Test Accuracy (PRISMA-DTA) statement [10]. The review protocol is registered on the international prospective register of systematic reviews (PROSPERO), CRD42021269563.

## Data sources and searches

The full strategy is listed in Supplementary Material 1. Searches were conducted on MEDLINE, EMBASE, the Cochrane library and Web of Science for studies published until September 2021. Bibliographies from eligible studies and systematic reviews were searched for additional relevant studies. Conference abstracts and pre-prints were excluded. A full description of data extraction is provided in Supplementary Material 2.

## Index test, reference standard and target condition

The target condition of the systematic review was the abnormality detected, for example intracranial haemorrhage. The AI model detecting the target condition was the index test. The radiological review was designated as the reference standard.

## Inclusion criteria

We included studies where an AI model predicted the presence of an abnormality using a CT or MRI examination. Only studies that validated AI models on test datasets that were separated from the training data temporally or geographically were included. Test datasets were required to have normal scans, scans with the target condition and scans with one or more non-target conditions, in order to be representative of clinical practice.

## Exclusion criteria

The motivation for the study was to review abnormality detection in first-line, clinical neuroimaging. Studies using second-line imaging exclusively (e.g. angiography, perfusion studies) were excluded. Psychiatric conditions were excluded if structural differences have only been shown in group-wise comparison studies (e.g. schizophrenia, autism spectrum disorder); all conditions that had a structural correlate often seen at the individual level were included (e.g. Alzheimer’s disease). Studies testing exclusively on paediatric populations were excluded. Studies not published in a peer-reviewed journal or without an English language translation were excluded [11].

## Data analysis

We used the QUality Assessment of Diagnostic Accuracy Studies-2 (QUADAS-2) tool [12], tailored to the review question incorporating items from the Checklist for Artificial Intelligence in Medical Imaging (CLAIM) [13]; modified signalling questions are presented in Supplementary Material 3. The unit of analysis was the patient undergoing a CT or MRI examination. The primary outcome was diagnostic test accuracy. Secondary outcomes assessed whether the AI model had been applied in a clinical pathway, and if so, the associated performance metrics.

### *Meta-analysis*

Meta-analysis was performed when four or more studies evaluated a given target condition within a specific modality [14]. Studies investigating the detection of intracranial haemorrhage on CT scans were the only subgroup of sufficient number and homogeneity to allow inclusion for meta-analysis. A bivariate random-effects model was used for meta-analysis (further details in Supplementary Material 2).

# RESULTS

## Characteristics of included studies

Database searches resulted in 42,870 unique results, of which 5,734 potentially eligible full texts were assessed (Supplementary Figure 1). Our criteria for clinically representative test datasets were not met in 1,239 studies which were excluded. Additionally, we excluded 218 studies for using internal validation only. Only 16 studies were of sufficient scientific rigour to be eligible for inclusion. The test datasets from the 16 studies comprised of 26,164 patients in total, however, the total number of patients in the training datasets could not be calculated as some commercial studies did not publish this data. Fourteen studies (14/16, 88%) used supervised convolutional neural networks (CNNs) to classify scans as normal or abnormal, with a variety of different model architectures. Five studies (5/16, 31%) demonstrated the accuracy of commercially available AI models (from three AI vendors:

Qure.ai (Mumbai, India); Aidoc (Tel Aviv, Israel); Avicenna.ai (La Ciotat, France) [15–20]). The largest subgroup of studies (11/16, 69%) employed CNNs to detect intracranial haemorrhage using CT [15,16,26,17–19,21–25]. Other studies used CT and MRI to detect other single non-haemorrhage pathology (2/16, 13%) [15,27], or multiple pathologies (4/16, 25%) [27–30]. The characteristics of each included study is summarised in Supplementary Material 5.

## Assessment of risk of bias

The risk of bias evaluation for each study using the QUADAS-2 tool is summarised in Figure 1. A high risk of bias in at least one domain was shown in 15/16 (94%) of studies. The modified signalling questions used for assessing each study, and their explanations are in Supplementary Materials 3 and 6 respectively.

The following were the commonest sources of bias. Eight studies (8/16, 50%) assessed AI model performance in laboratory conditions only ("analytical validation" [31]) [15,19–23,27,29]. In contrast, four studies (4/16, 25%) placed the AI model within the clinical pathway ("clinical validation") [16–18,25], which more closely resembles a "real world" environment and therefore the intended applicability. Seven studies (7/16, 44%) used temporal validation alone, and therefore had a high risk of bias for patient selection as there is limited assessment of generalisability [20,23–25,28–30], compared to 9/16 (56%) studies where AI models were externally validated on test data from other institutions [15–19,21,22,26,27]. Studies that used fewer than two radiologists to assess the images of a scan for their reference standard were considered at high risk of bias, as individual radiologists do not have perfect accuracy - five studies (5/16, 31%) were therefore considered to have high risk of bias as only the clinical report was reviewed [16,25–27,30]. One study had a high risk of bias as the reference standard was informed by the output of the AI model (the index test) [19]; this study was therefore excluded from the meta-analysis.

**Fig. 1 Summary of the QUADAS-2 risk of bias assessment**

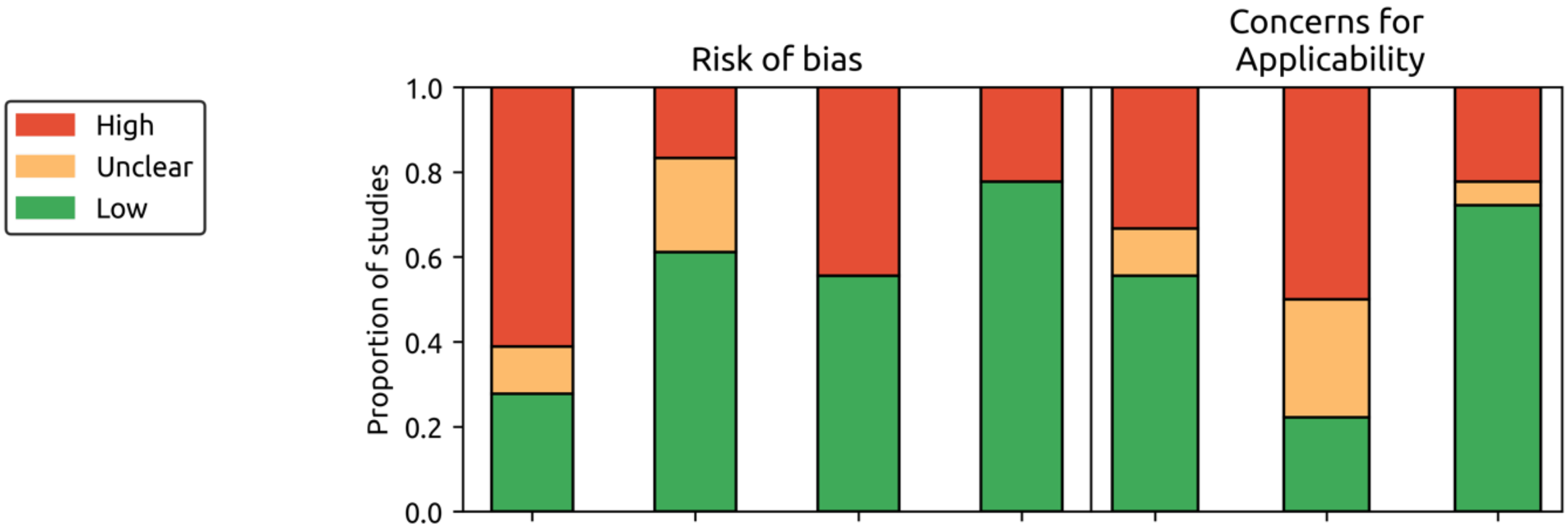

| Study | Risk of bias: Patient selection | Risk of bias: Index test | Risk of bias: Reference standard | Risk of bias: Flow and timing | Concerns for Applicability: Patient selection | Concerns for Applicability: Index test | Concerns for Applicability: Reference standard |
|---|---|---|---|---|---|---|---|
| Arbabshirani (2018) | High | Low | High | High | Low | Low | High |
| Buls (2021) | Low | Low | Low | High | Low | Low | Low |
| Chang (2018) | High | Unclear | Low | Low | Low | Unclear | Low |
| Chilamkurthy (2018) | High | High | Low | Low | High | High | Low |
| Finck (2021) | High | Low | Low | Low | Low | Unclear | Unclear |
| Gauriau (2021) | High | Low | Low | Low | Low | High | High |
| Ginat (2020) | Low | Low | Low | Low | Low | Low | Low |
| Ginat (2021) | Low | Low | High | Low | Low | Low | Low |
| Kuo (2019) | High | High | Low | High | High | High | Low |
| McLouth (2021) | High | Low | Low | Low | High | High | Low |
| Monteiro (2020) | High | High | Low | Low | High | High | Low |
| Nael (2021) ICH, infarct | Unclear | Low | High | Low | Unclear | High | Low |
| Nael (2021), all pathology | Unclear | Low | High | Low | Unclear | High | High |
| Prevedello (2017), multiple | High | Unclear | High | Low | Low | Unclear | Low |
| Prevedello (2017), infarct | High | Unclear | High | High | High | Unclear | Low |
| Salehinejad (2021) | Low | Low | High | Low | Low | Unclear | High |
| Voter (2021) | Low | Low | High | Low | Low | High | Low |
| Wang (2021) | High | Unclear | Low | Low | High | High | Low |

## Analysis

The primary outcome for each study was diagnostic test accuracy, which is summarised in Table 1. Studies varied greatly in accuracy performance (sensitivity range: 0.70 – 1.00, specificity range: 0.51 – 1.00), but the highest accuracies were typically seen for intracranial haemorrhage detection using CT (Supplementary Figure 2). Only two studies validated AI that used MRI; two AI models that detected any pathology using MRI had modest accuracies (sensitivity range: 0.78 – 1.00, specificity range: 0.65 – 0.80), compared to single pathology performance (Supplementary Figure 3).

*Diagnostic test accuracy of radiologists compared to AI*

The performance of AI models against individual radiologists under laboratory conditions are available for four studies, summarised in Figure 2. A full description of these studies can be found in Supplementary Material 7.

**Fig. 2 Diagnostic test accuracy of algorithms and comparators (single radiologists) in precision-recall space, for intracranial haemorrhage detection in CT imaging, when adjusted for an intracranial haemorrhage prevalence of 10%. An ideal classifier would be at the top-right corner at (1, 1). The size of each marker is proportional to the size of the test dataset. In studies where there was more than one operating point, we chose the operating point with the highest specificity. CQ500 = CQ500 external test dataset. Qure.ai, Aidoc and Avicenna.ai are commercial vendors for AI products.**

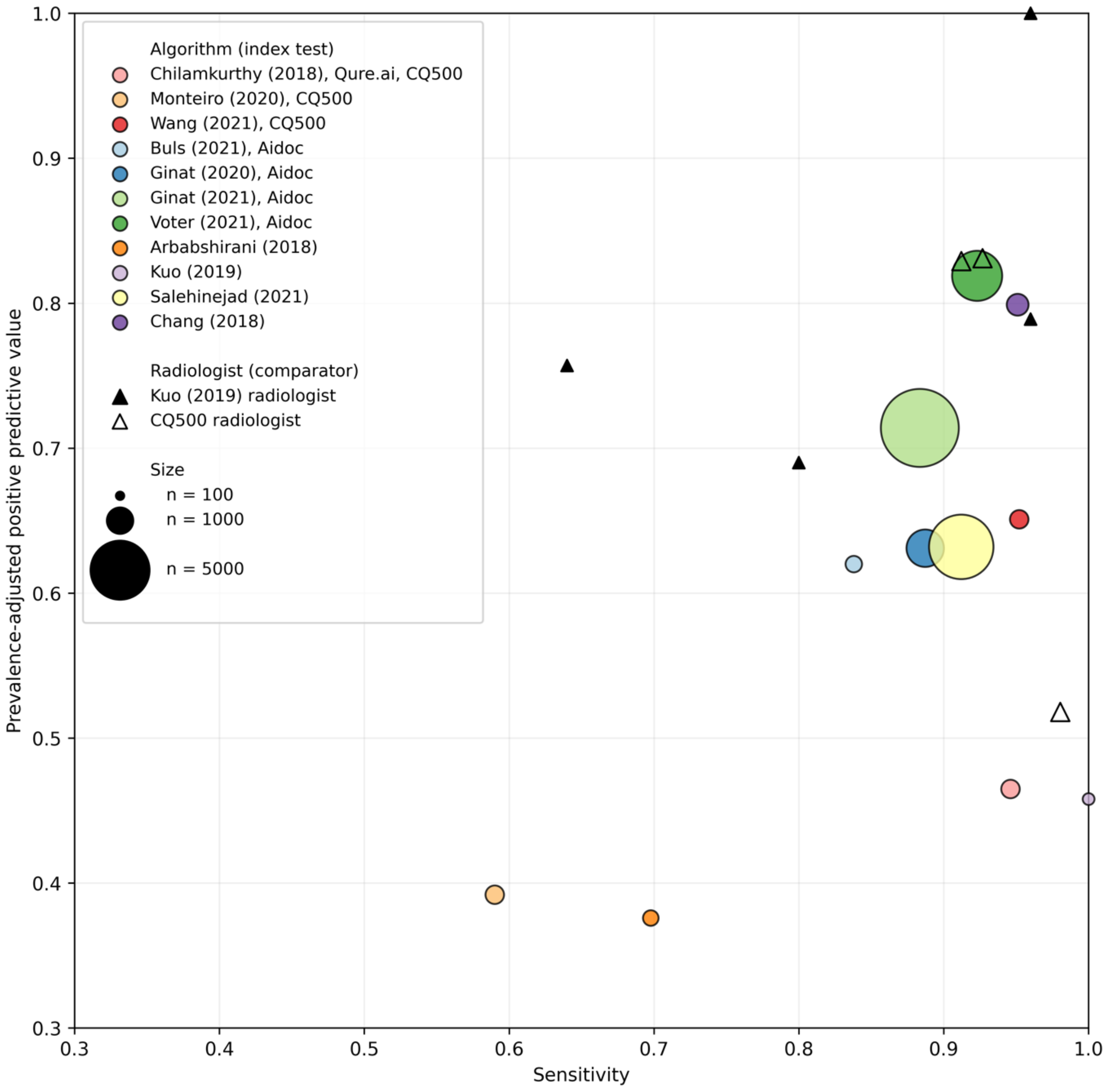

*Clinical implementation*

All 3/16 (19%) studies that also investigated clinical implementation performance, assessed the detection of intracranial haemorrhage using CT. Two (2/16, 13%) studies placed the AI model at the start of the clinical pathway before radiologist interpretation (pre-read triage) [16,25], and 2/16 (13%) at the end after radiologist interpretation (post read) [25,26].

In the two studies where AI had been applied for pre-read triage, one showed a reduction in the time-to-report for non-urgent examinations which the AI flagged as abnormal, from a median time of 512 minutes to 19 minutes [25]. The other demonstrated significant reductions in the mean time-to-report for flagged examinations for outpatients (674 to 70 minutes, $p < 0{\cdot}001$), inpatients (390 to 352 minutes, $p = 0{\cdot}002$), but not emergency cases ($p = 0{\cdot}37$), or an undefined “other” class ($p = 0{\cdot}25$) [16]. Importantly, neither study examined the extent and potential harms of delaying non-flagged studies, particularly AI false negatives.

In two studies (2/16, 13%), AI had been applied as a second reader after radiologist interpretation and discrepancies between radiologists and AI were examined [25,26]. AI was able to identify 4/347 (1·2%) and 2/5965 (0·03%) of intracranial haemorrhages that radiologists had missed (radiologist false negatives) respectively. If implemented, both studies estimated that the radiologist would be alerted that there was a discrepancy between them and the AI model in 10% (34/347) and 5% (313/5965) of cases respectively, and 9 and 157 re-reviews would be required for 1 change in report respectively [25]. In the second study, radiologist-positive and AI-negative discrepancies were also examined (59/5965, 1%) and found these were all AI false negatives - AI was unable to identify any radiologist overcalls (radiologist false positives). If AI were to be implemented to identify overcalls (radiologist false positives) as well as misses (radiologist false negatives), the radiologist would be alerted in 6% (372/5965) of cases and 186 re-reviews would be required for 1 change in a report [26].

*Meta-analysis*

Six (6/16, 38%) studies were unsuitable for meta-analysis. We excluded five studies (5/16, 31%) with heterogeneous target conditions (one study detected acute intracranial haemorrhage only without non-acute cases) or modality [20,27–30]. One study (1/16, 6%) was excluded due to the fundamental methodological flaw of having a circular reference standard as described above [19]. The remaining subgroup of studies were those detecting intracranial haemorrhage using CT and applying CNNs, and consisted of ten studies (10/16, 63%) [15–18,21–26]; we included these studies for meta-analysis. Forest plots of sensitivity and specificity (Figure 3) graphically showed a high level of heterogeneity. Chi-squared tests also demonstrated significant heterogeneity in sensitivity and specificity (p-values both < 0·001). The pooled sensitivity for intracranial haemorrhage detection in CT = 0·901 (95% CI 0·853 to 0·935), and the pooled specificity = 0·903 (95% CI 0·826 to 0·948). The derived pooled measures of balanced accuracy = 0·931 (95% CI 0·889- 0·957); positive likelihood ratio = 26·7 (95% CI 15·8 - 42·3); negative likelihood ratio = 0·106 (95% CI 0·0471- 0·199); and diagnostic odds ratio = 280·0 (95% CI 128·0- 533·0).

**Fig. 3 Forest plots demonstrating individual studies' sensitivities and specificities**

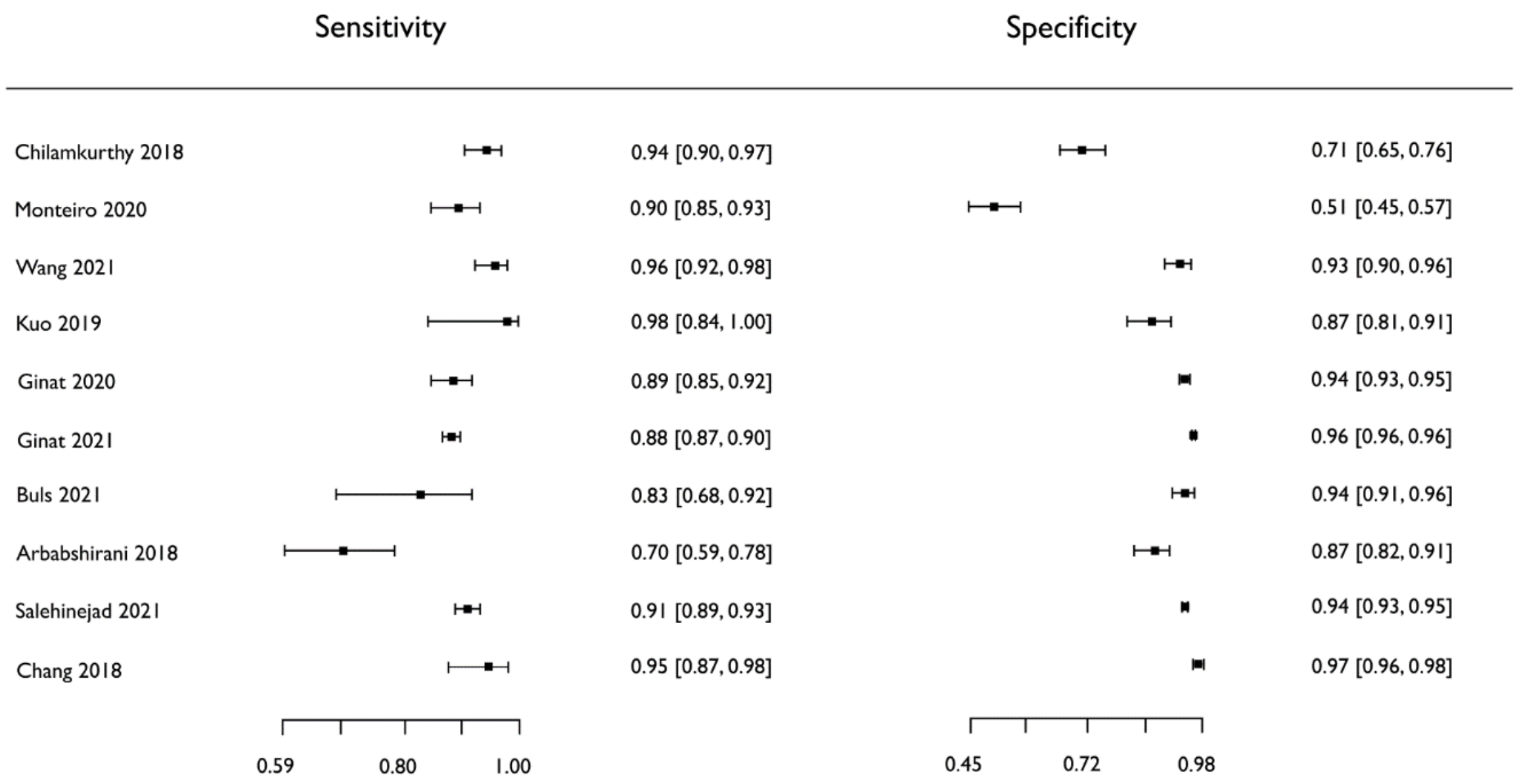

Heterogeneity was investigated using meta-regression, which compared the pooled sensitivities and specificities of two subsets of the studies: three studies where different AI models were applied on the same test dataset (CQ500) [15,21,22], and three studies where the same AI model (Aidoc) was applied on different test datasets [16–18]. Using the Aidoc subset as a baseline, the CQ500 subset had higher pooled sensitivity (p = 0·008) and lower pooled specificity (p = 0·004), implying that AI model type and patient make-up in the test dataset contributed to the heterogeneity observed.

Individual study ROC point estimates resulted in a summary ROC (SROC) curve (Figure 4), for which the summary ROC-AUC = 0·948.

**Fig. 4 Summary receiver operating characteristic (SROC) curve for intracranial haemorrhage detection in CT imaging**

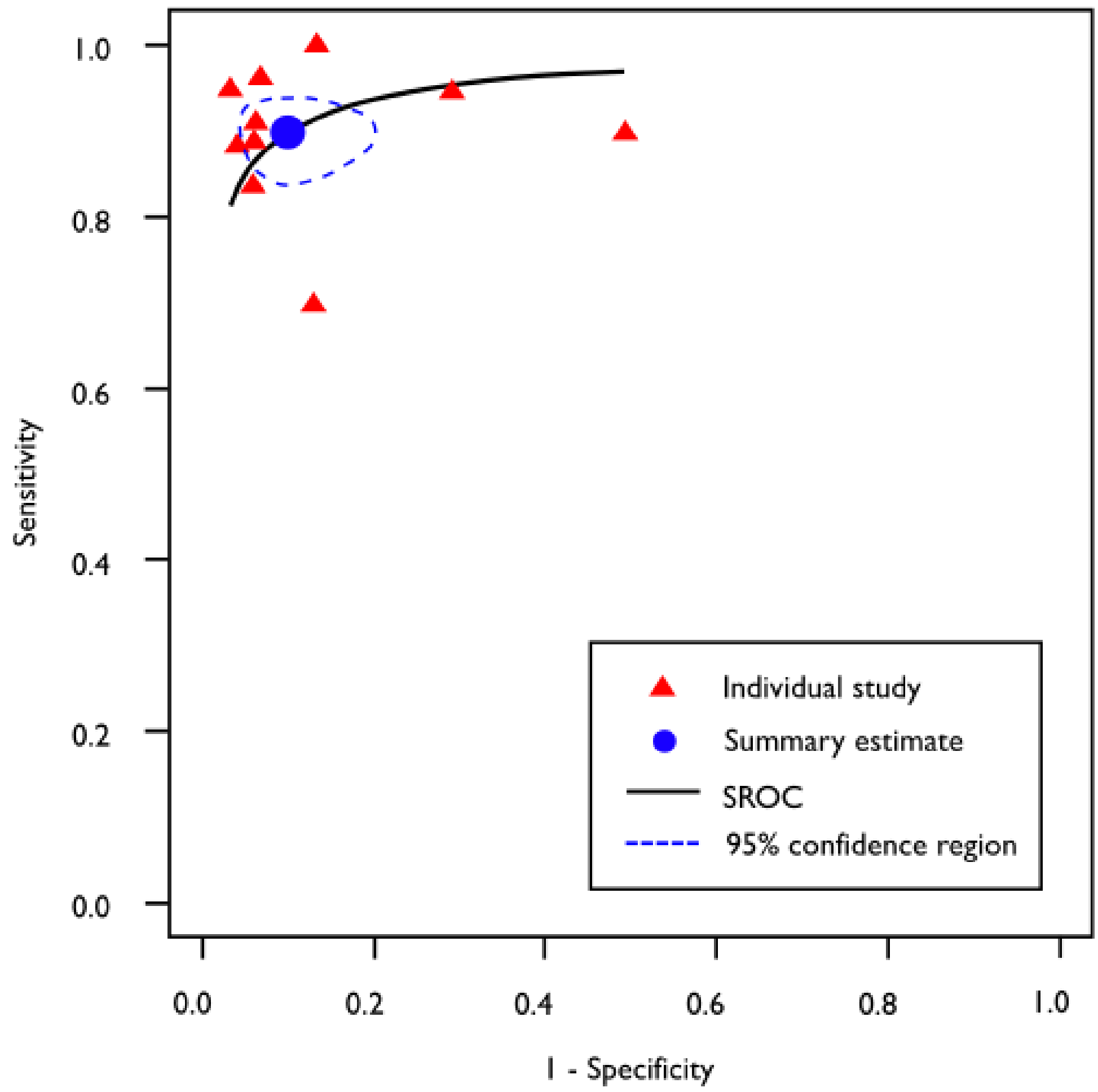

# DISCUSSION

## Summary

This study aimed to determine the diagnostic accuracy of AI systems used to identify abnormalities in first-line neuroimaging tasks. Any productivity gains in such tasks are important as they are high volume and performed in almost all hospitals. To ensure our analyses were only focused on those studies that were not compromised by unrepresentative datasets or inadequate validation methodology, we excluded many studies (1,239) that did not validate the AI model using datasets from representative clinical cohorts and many (218) for validating without temporal or external validation. Only 16 studies were of sufficient rigour to be eligible for inclusion, however, even for these studies the overall methodological quality remained low with a high risk of bias in 94% of studies. Furthermore, most included studies were retrospective, with only four studies validating their AI models in clinical environments prospectively in real time (i.e. clinical validation).

For CT imaging studies, a subgroup of ten AI models used to detect intracranial haemorrhage using CNNs, had a pooled sensitivity and specificity of 0·90, with a summary ROC-AUC of 0·95. Meta-regression suggested that differences in the AI model development and patient selection contributed to the significant heterogeneity observed in both pooled measures. Four CT imaging studies allowed for direct comparison between AI models with radiologists under laboratory conditions – further discussion is provided in Supplementary Material 7.

For MRI, only two studies were included. Both studies validated AI models that detect all pathologies. Together with a third study that used CT, a limitation of the three AI models that detect all pathologies is that findings seen in healthy ageing such as small vessel disease and age-commensurate atrophy are considered abnormal – this is reflected in the high prevalence of what was assigned as pathological in their test datasets (64 – 81%) [27–29]. AI that overcalls all elderly patients as abnormal raises concerns for applicability in clinical practice.

There were only three clinical implementation studies where AI was placed within the clinical pathway, as pre-read triage and for post-read discrepancy identification. No study demonstrated a downstream clinical or health economic benefit.

## Strengths and limitations

A strength of this study was that the search strategy was sensitive [32,33]. This allowed for the identification of a wide range of studies included in this review, many of which were missing in other systematic reviews for the general use of AI in neuroimaging [34–37]. We also included all AI methods, not just those limited to deep learning. Whilst broad inclusion is a study strength, it is also conceivable that summary performance accuracy might be diminished by the inclusion of older AI models. However, older AI models barely contributed to our results as 88% of all eligible studies, and 100% of studies in the meta-analysis subgroup used CNNs.

Another strength, unique to this study, was that the inclusion criteria were designed to only include studies where outcome metrics would have a reasonable chance of generalising to first-line neuroimaging in routine clinical practice. Therefore, the diagnostic test accuracies presented here are plausibly more generalisable than if less stringent inclusion criteria were used. Specifically, we first excluded studies that did not validate AI on temporally distinct or external test datasets. Second, we excluded studies that did not test on representative patient cohorts (which as a minimum standard required normal brains, the target condition and at least one non-target condition). As a result we excluded studies that validated AI models on test datasets that contained the target condition and healthy controls only, which does not reflect the “real world”; we note that almost all ischaemic stroke detection studies were therefore excluded [38].

A limitation is that meta-analysis was only suitable for one subgroup where there were sufficient homogenous studies using the same imaging modality, target condition and AI model type. Another limitation is that no formal assessment of publication bias was undertaken, however, it is unlikely that our overall conclusions would change if studies with poorer AI model performances had been published.

## Strategies for implementing AI into clinical pathways

The standalone diagnostic accuracy of AI to detect abnormalities has been demonstrated to be high, particularly for intracranial haemorrhage in CT imaging. There was insufficient evidence, however, to suggest where such AI would be most useful in the clinical pathway. This included those being marketed commercially (Supplementary Material 8).

Both studies that investigated intracranial haemorrhage AI detectors for pre-read CT worklist triage found that the greatest reduction in reporting time was for outpatient examinations when compared to emergency or inpatient examinations [16,25]. There was insufficient evidence however, from these and any other studies regarding the downstream clinical benefit and cost effectiveness of AI implementation. Inaccurate AI models in a pre-read triage setting would systemically increase the time to report AI false negative examinations as these would be put at the back of the reporting queue; it was unclear from both studies whether AI false negatives were significantly delayed and the extent of harm, if any, that was associated with this delay. For pre-read triage, it is also unknown whether knowing that AI puts flagged examinations to the front of the queue could have long-term consequences on radiologist performance. There is a similar question regarding AI intended as a second reader which may unintentionally affect the behaviour of radiologists; for example, the implementation of automated computer-assisted-diagnosis (CAD) tools in mammography, to be used as a reader aid during radiologist interpretation, has previously been shown to reduce radiologist sensitivity [39] and overall accuracy [40].

One advantage of a post-read implementation is that radiologists are initially blinded to the AI decision. In a post-read setting, AI models could be used to flag discrepancies to determine potential radiologist “misses” or “overcalls” and allow a re-review [41]. Regardless of strategy, an inaccurate AI model in this setting would create a high burden on radiologist time. In the two studies that investigated discrepancies, there appeared to be low additive diagnostic yield associated with a high rate of re-review. Therefore, further studies will be necessary to understand the cost-effectiveness of such post-read strategies.

## Conclusions

We have analysed the evidence and presented the diagnostic performance of the current state-of-the-art AI detection models that can be applied to first-line neuroimaging. Such tasks are important as they are high volume and performed in almost all hospitals and offer considerable potential for the necessary productivity gains required in the 21st century. If the intended use of AI detection models is as a tool to improve radiologist efficiency rather than a replacement for radiologists, AI may be clinically useful even if the accuracy shown in our meta-analysis remains lower than that of radiologists for the task of intracranial haemorrhage detection. However, at present, there is insufficient evidence to recommend implementation of AI for abnormality detection, including haemorrhage detection, into any part of the clinical pathway. Importantly, the clinical and health economic benefits are currently unproven. For now, future research efforts should aim to minimise bias and demonstrate analytical validation through well-designed studies using clinically representative external test datasets which can unequivocally prove high performance accuracy and good generalisability [42,43,44]. Following this, clinical trials will be required to confirm the performance findings in the "real world" and determine whether the clinical benefits of implementing AI in the clinical pathway outweigh the potential harm to patients. In addition to clinical validation, such trials could include health economic analyses to determine the costs incurred and benefits obtained within the wider healthcare system.

# TABLES

**Table 1: Diagnostic test accuracy for included studies. Developers can choose different "operating points" which allows AI models to favour either sensitivity or specificity. The shaded cells indicate which values had to be calculated from the published data.**

| Study | Modality, Target | Training set (n) | Test set (n) | P (n, %) | Validation (test set separation, and if clinically validated) | TP | FN | FP | TN | ROC-AUC (95 CI) | Sensitivity | Specificity | PPV | Prevalence adjusted-PPV |
|---|---|---|---|---|---|---|---|---|---|---|---|---|---|---|
| Chilamkurthy (2018), Qure.ai, high sensitivity operating point | CT, ICH | 4304 (165809 slices) | 491 (CQ500 dataset) | 205 (42%) | Geographical | 194 | 11 | 83 | 203 | 0.942 (0.919 - 0.965) | 0.946 | 0.710 | 0.700 | 0.266 |
| Chilamkurthy (2018), Qure.ai, high specificity operating point | | | | | | 168 | 37 | 30 | 256 | | 0.820 | 0.895 | 0.848 | 0.465 |
| Chilamkurthy (2018), Radiologist comparators | | N/A | | | N/A | 201 | 4 | 29 | 257 | N/A | 0.980 | 0.899 | 0.874 | 0.518 |
| | | | | | | 190 | 15 | 6 | 280 | | 0.927 | 0.979 | 0.969 | 0.831 |
| | | | | | | 187 | 18 | 6 | 280 | | 0.912 | 0.979 | 0.969 | 0.829 |
| Monteiro (2020), high sensitivity operating point | | 655 | | | Geographical | 184 | 21 | 140 | 145 | 0.83 (0.79 - 0.87) | 0.898 | 0.509 | 0.568 | 0.169 |
| Monteiro (2020), high specificity operating point | | | | | | 121 | 84 | 29 | 256 | | 0.590 | 0.898 | 0.807 | 0.392 |
| Wang (2021), high sensitivity operating point | | 19530 (674258 slices) | | | Geographical | 198 | 8 | 19 | 266 | 0.985 (0.977 - 0.993) | 0.961 | 0.933 | 0.912 | 0.616 |
| Wang (2021), high specificity operating point | | | | | | 197 | 10 | 16 | 266 | | 0.952 | 0.943 | 0.925 | 0.651 |
| Kuo (2019) | | 4396 | 200 | 25 (13%) | Temporal | 25 | 0 | 23 | 152 | 0.991 (0.985 - 0.997) | 1.00 | 0.87 | 0.521 | 0.458 |
| Kuo (2019), Radiologist comparators | | N/A | | | | 24 | 1 | 5 | 170 | N/A | 0.96 | 0.97 | 0.828 | 0.789 |
| | | | | | | 24 | 1 | 0 | 175 | | 0.96 | 1.00 | 1.000 | 1.000 |
| | | | | | | 16 | 9 | 4 | 171 | | 0.63 | 0.98 | 0.800 | 0.757 |
| | | | | | | 20 | 5 | 7 | 168 | | 0.79 | 0.96 | 0.741 | 0.690 |
| Ginat (2020), Aidoc | | ~50000 | 2011 | 373 (19%) | Geographical, clinical | 275 | 35 | 98 | 1603 | | 0.887 | 0.942 | 0.737 | 0.631 |
| Ginat (2021), Aidoc | | | 8723 | 1760 (20%) | Geographical, clinical | 1555 | 205 | 274 | 6689 | | 0.884 | 0.961 | 0.850 | 0.714 |
| Buls (2021), Aidoc | | | 388 | 37 (10%) | Geographical, clinical | 31 | 6 | 20 | 331 | | 0.838 | 0.943 | 0.608 | 0.620 |
| Voter (2021), Aidoc | | | 3605 | 349 (10%) | Geographical | 322 | 27 | 74 | 3182 | | 0.923 | 0.977 | 0.813 | 0.819 |
| Arbabshirani (2018) | | 24882 | 347 | 86 (25%) | Temporal, clinical | 60 | 26 | 34 | 230 | | 0.698 | 0.871 | 0.638 | 0.376 |
| Chang (2018) | | 10159 | 682 | 82 (12%) | Temporal | 78 | 4 | 16 | 584 | 0.981 | 0.951 | 0.973 | 0.829 | 0.799 |
| Salehinejad (2021) | | 21784 | 5965 | 674 (11%) | Geographical | 615 | 59 | 313 | 4978 | 0.954 | 0.912 | 0.941 | 0.663 | 0.819 |
| McLouth (2021), Avicenna.ai | CT, acute ICH | 8994 | 814 | 255 (31%) | Temporal | 233 | 22 | 14 | 545 | | 0.914 | 0.975 | 0.943 | 0.802 |
| Prevedello (2017) | CT, ICH, mass effect, acute hydrocephalus | 197 | 80 | 50 (63%) | Temporal | 45 | 5 | 12 | 68 | 0.91 | 0.900 | 0.850 | 0.789 | |
| Prevedello (2017) | CT, infarct | 57 | 49 | 21 (43%) | Temporal | 13 | 8 | 1 | 27 | 0.81 | 0.619 | 0.964 | 0.929 | |
| Chilamkurthy (2018), Qure | CT, skull fracture | 290055 | 491 | 39 (8%) | Geographical | 37 | 2 | 61 | 391 | 0.962 (0.920 - 1.00) | 0.949 | 0.865 | 0.378 | |
| Chilamkurthy (2018), Qure | CT, mass effect | 290055 | 491 | 115 (26%) | Geographical | 115 | 12 | 23 | 104 | 0·922 (0·888 - 0·955) | 0.906 | 0.819 | 0.833 | |
| Finck (2021), 'uncertain' label considered abnormal | CT, any abnormality | 191 | 248 | 178 (72%) | Temporal | 178 | 0 | 44 | 26 | | 1.000 | 0.371 | 0.802 | |
| Nael (2021) | MR, any abnormality | 9845 | 1072 | 867 (81%) | Geographical | 691 | 175 | 41 | 164 | 0.88 | 0.80 | 0.80 | 0.94 | |
| | ICH | 9845 | 1072 | 78 (7%) | Geographical | 56 | 22 | 119 | 875 | 0.83 | 0.72 | 0.88 | 0.32 | |
| | Infarct | 9845 | 1072 | 287 (27%) | Geographical | 258 | 29 | 23 | 762 | 0.97 | 0.90 | 0.97 | 0.92 | |
| | Mass effect | 9845 | 1072 | 31 (3%) | Geographical | 25 | 6 | 198 | 843 | 0.87 | 0.81 | 0.81 | 0.12 | |
| Gauriau (2021) | MR, any abnormality | 2741 | 1489 | 960 (64%) | Temporal | 742 | 218 | 188 | 346 | 0.80 (0.77 - 0.82) | 0.773 | 0.648 | 0.798 | |

P (positive) = number of examinations with target pathology in test dataset, TP = true positives, FN = false negatives, FP = false positives, TN = true negatives, ROC-AUC = area under receiver operating characteristic curve, PPV = positive predictive value, ICH = intracranial haemorrhage. Qure.ai, Aidoc and Avicenna.ai are commercial vendors for AI products.

# SUPPLEMENTARY FIGURES WITH FIGURE LEGENDS

Supplementary Figure 1: PRISMA flow diagram

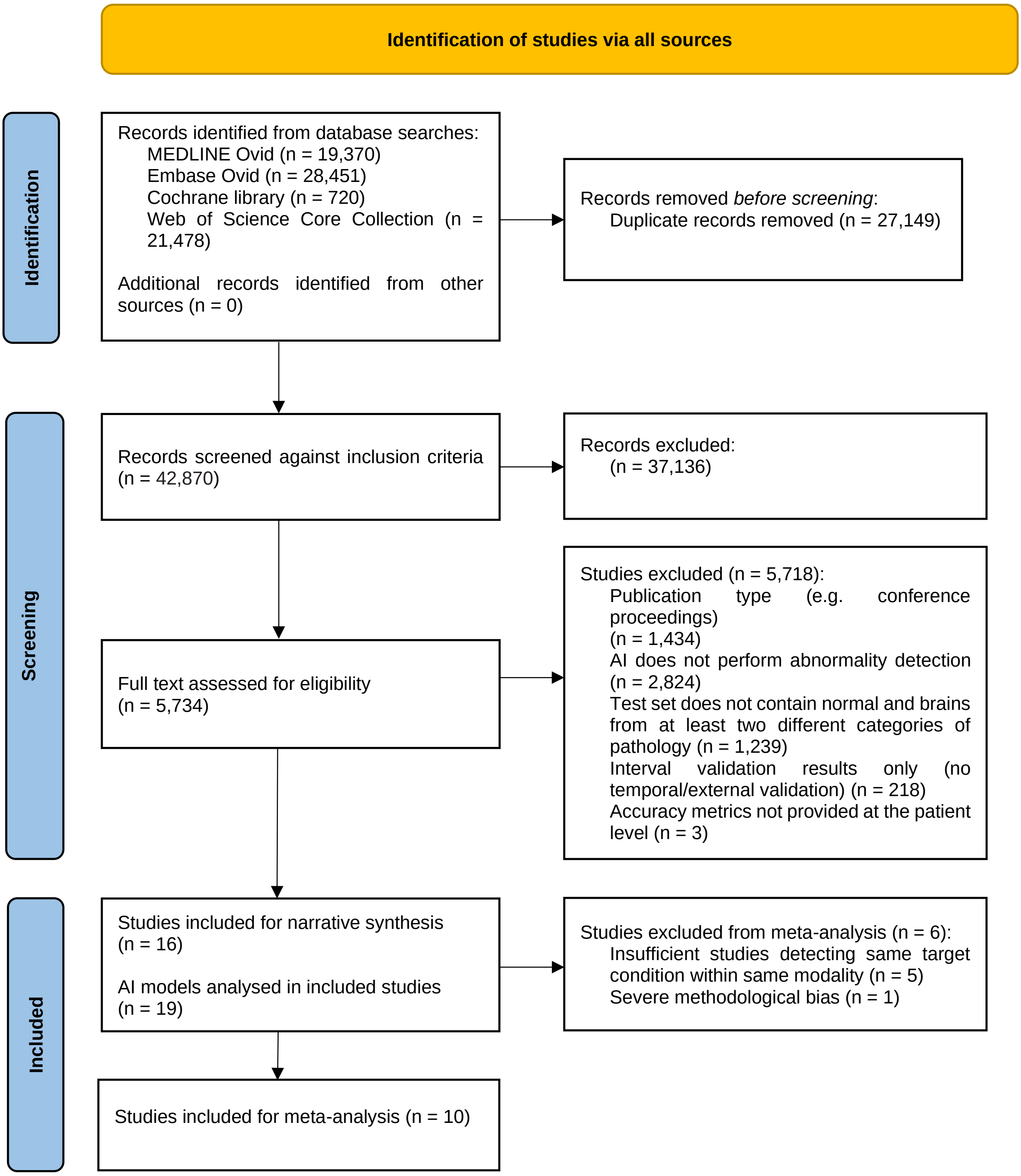

Supplementary Figure 2: Diagnostic test accuracy of AI models in CT imaging in receiver operating characteristic (ROC) space, compared with individual radiologists. An ideal classifier would be at the top-left corner at (0, 1). The size of each marker is proportional to the size of the test dataset. In studies where there was more than one operating point, we chose the operating point with the highest sensitivity. ICH = intracranial haemorrhage. CQ500 = CQ500 external test dataset. Qure.ai, Aidoc and Avicenna.ai are commercial vendors for AI products.

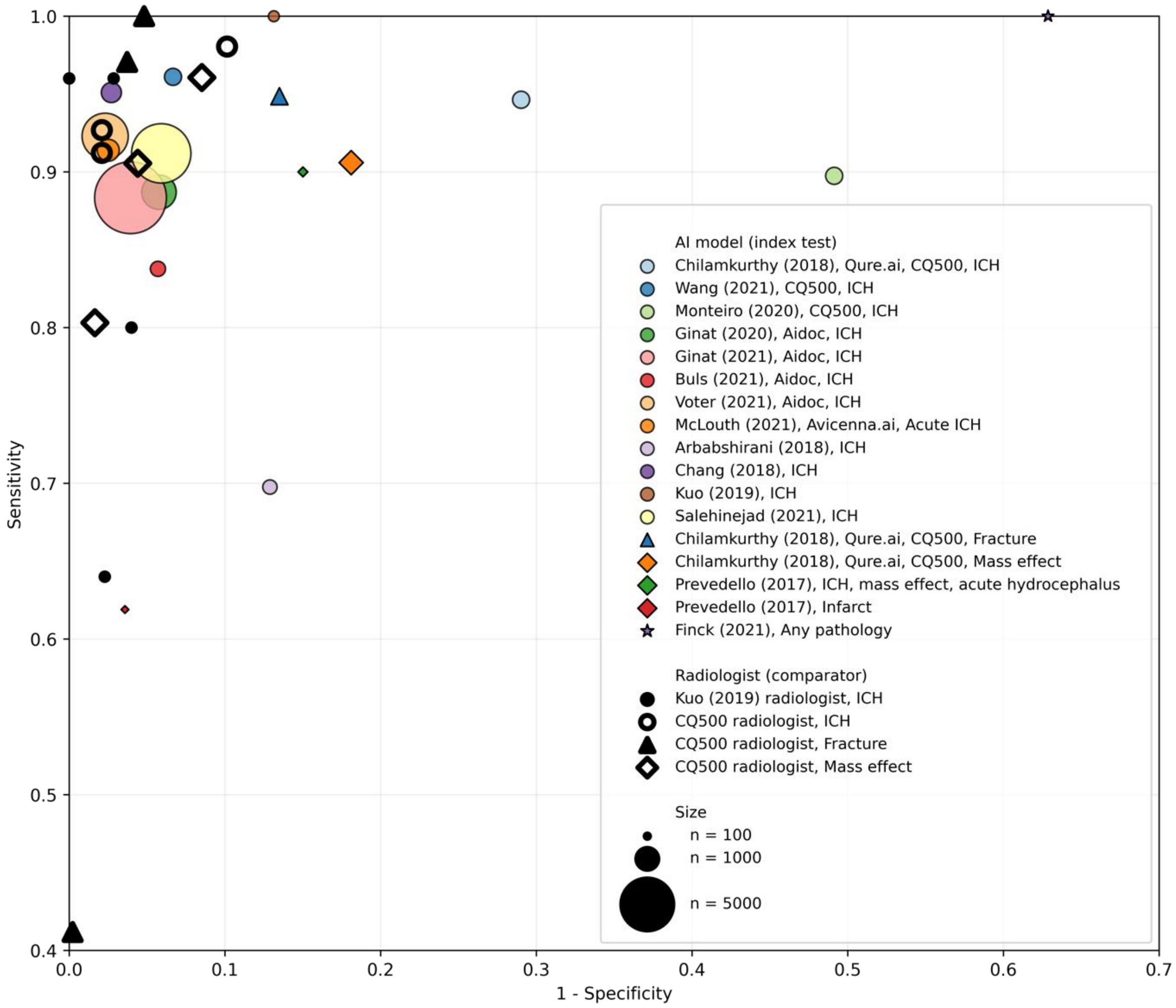

Supplementary Figure 3: Diagnostic test accuracy of AI models in MRI in ROC space. The size of each marker is proportional to the size of the test dataset. An ideal classifier would be at the top-left corner at (0, 1).

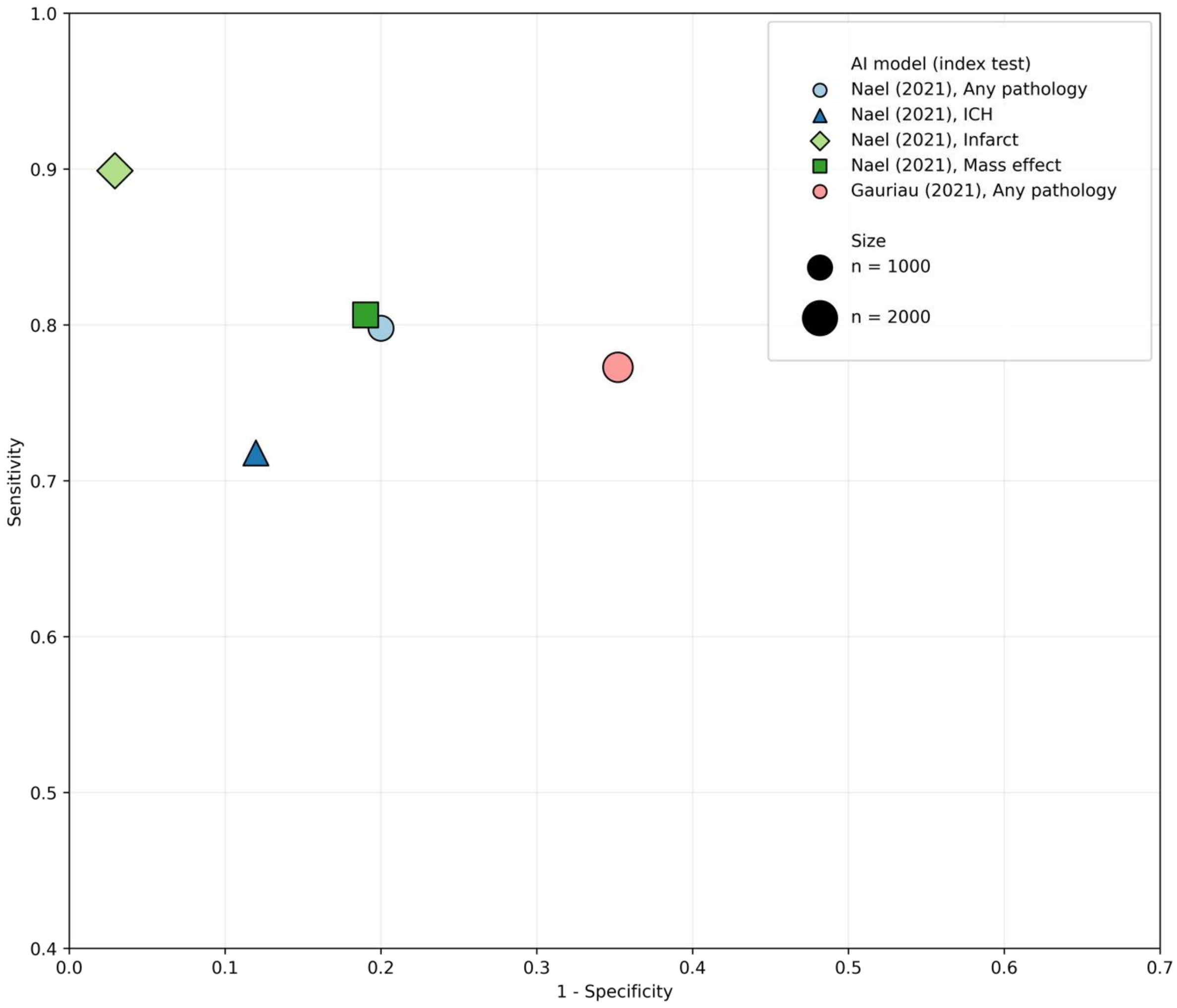

**Supplementary Material: Artificial intelligence for abnormality detection in high volume neuroimaging: a systematic review and meta-analysis**

## CONTENTS

## SUPPLEMENTARY MATERIAL 1: SEARCH STRATEGY

MEDLINE (OVID)

1. exp Magnetic Resonance Imaging/
2. (mri or mr or (magnetic adj1 resonance)).ab,ti.
3. exp Tomography, X-Ray Computed/
4. (ct or (comput* adj1 tomograph*)).ab,ti.
5. (cat adj3 (scan* or imag* or stud*)).ab,ti.
6. neuroimag*.ti,ab.
7. 1 or 2 or 3 or 4 or 5 or 6
8. (brain or head or skull or cerebral or intracerebral or cerebrum or cranial or intracranial or cranium).ab,ti.
9. (detect* or classif* or identif* or diagnos* or predict* or decis* or decid*).ti,ab.
10. 7 and 8 and 9
11. exp Diagnosis, Computer-Assisted/ or exp Algorithms/ or exp Artificial Intelligence/ or exp Machine Learning/ or exp Neural Networks, Computer/ or exp Pattern Recognition, Automated/
12. ((artificial adj1 intelligence) or ((deep or machine) adj1 learning)).mp.
13. algorithm*.ab,ti,kw,kf. or automat*.ab,ti. or radiomic*.mp. or (comput* adj3 (aid* or assist* or vision*)).ti,ab.
14. ((supervised or unsupervised or (semi adj1 supervised) or deep hybrid or cluster* or bayes* or gauss*) adj3 (learning or model* or net* or algo*)).mp.
15. ((feature adj3 (engineer* or select* or extract* or learn*)) or hyperparameter).mp. [mp=title, abstract, original title, name of substance word, subject heading word, floating sub-heading word, keyword heading word, organism supplementary concept word, protocol supplementary concept word, rare disease supplementary concept word, unique identifier, synonyms]
16. (((neural or conv*) adj1 (net* or learn* or model*)) or CNN or convnet or RNN, or long short-term memory or lstm or gate* recurrent unit or gru or boltzmann machine or deep belief net* or spatial transformer net* or sum product network).mp.
17. ((ensemble or transfer or zero shot or reinforcement or dictionary) adj1 (learning or model* or net* or algo*)).mp.
18. (vector machine or SVM or ((classification or regression or probability or decision) adj1 tree*) or random forest).mp.
19. (generative model* or autoencod* or aae or vae or cae or dae or sdae or gan or generative adversarial).mp.
20. (pca or principal component analysis or (k adj1 means) or (nearest adj1 neighbo?r) or knn or (fuzzy adj3 logi*) or isolation forest or hidden markov model or association rule* or feature bag* or score normali$ation).mp.
21. 11 or 12 or 13 or 14 or 15 or 16 or 17 or 18 or 19 or 20
22. 10 and 21

EMBASE (OVID)

1. exp nuclear magnetic resonance imaging/
2. (mri or mr or (magnetic adj1 resonance)).ab,ti.
3. computer assisted tomography/
4. (ct or (comput* adj1 tomograph*)).ab,ti.
5. (cat adj3 (scan* or imag* or stud*)).ab,ti.
6. neuroimag*.ti,ab.
7. 1 or 2 or 3 or 4 or 5 or 6
8. (brain or head or skull or cerebral or intracerebral or cerebrum or cranial or intracranial or cranium).ab,ti.
9. (detect* or classif* or identif* or diagnos* or predict* or decision or decis* or decid*).ti,ab.
10. 7 and 8 and 9
11. computer assisted diagnosis/ or exp algorithm/ or exp artificial intelligence/ or exp machine learning/
12. ((artificial adj1 intelligence) or ((deep or machine) adj1 learning)).mp.
13. (algorithm* or automat*).ab,ti. or radiomic*.mp. or (comput* adj3 (aid* or assist* or vision*)).ti,ab.
14. ((supervised or unsupervised or (semi adj1 supervised) or deep hybrid or cluster* or bayes* or gauss*) adj3 (learning or model* or net* or algo*)).mp.
15. ((feature adj3 (engineer* or select* or extract* or learn*)) or hyperparameter).mp.

| |
|---|
| 16. (((neural or conv*) adj1 (net* or learn* or model*)) or CNN or convnet or RNN, or long short-term memory or lstm or gate* recurrent unit or gru or boltzmann machine or deep belief net* or spatial transformer net* or sum product network).mp.<br>17. ((ensemble or transfer or zero shot or reinforcement or dictionary) adj1 (learning or model* or net* or algo*)).mp.<br>18. (vector machine or SVM or ((classification or regression or probability or decision) adj1 tree*) or random forest).mp.<br>19. (generative model* or autoencod* or aae or vae or cae or dae or sdae or gan or generative adversarial).mp.<br>20. (pca or principal component analysis or (k adj1 means) or (nearest adj1 neighbo?r) or knn or (fuzzy adj3 logi*) or isolation forest or hidden markov model or association rule* or feature bag* or score normali$ation).mp.<br>21. 11 or 12 or 13 or 14 or 15 or 16 or 17 or 18 or 19 or 20<br>22. 10 and 21 |
| Web of Science<br><br>1. TI =(mri or mr or (magnetic NEAR/0 resonance) ) OR AB=(mri or mr or (magnetic NEAR/0 resonance) )<br>2. TI=(ct or (comput* NEAR/0 tomograph*) ) OR AB=(ct or (comput* NEAR/0 tomograph*) )<br>3. TI=(cat NEAR/2 (scan* or imag* or stud*) ) OR AB=(cat NEAR/2 (scan* or imag* or stud*) )<br>4. TI=(neuroimag*) OR AB=(neuroimag*)<br>5. #4 OR #3 OR #2 OR #1<br>6. TI=(brain or head or skull or cerebral or intracerebral or cerebrum or cranial or intracranial or cranium) OR AB=(brain or head or skull or cerebral or intracerebral or cerebrum or cranial or intracranial or cranium)<br>7. TI=(detect* or classif* or identif* or diagnos* or predict* or decision or decid*) OR AB=(detect* or classif* or identif* or diagnos* or predict* or decision or decid*)<br>8. #7 AND #6 AND #5<br>9. TI=((artificial NEAR/0 intelligence) or ((deep or machine) NEAR/0 learning)) OR AB=((artificial NEAR/0 intelligence) or ((deep or machine) NEAR/0 learning))<br>10. TI=(algorithm* or automat* or radiomic* or (comput* NEAR/2 (aid* or assist* or vision*) )) OR AB=(algorithm* or automat* or radiomic* or (comput* NEAR/2 (aid* or assist* or vision*) ))<br>11. TI = (supervised NEAR/2 (learning or model* or net* or algo*) ) OR AB = (supervised NEAR/2 (learning or model* or net* or algo*) )<br>12. TI = (unsupervised NEAR/2 (learning or model* or net* or algo*) ) OR AB= (unsupervised NEAR/2 (learning or model* or net* or algo*) )<br>13. TI=(semi supervised NEAR/2 (learning or model* or net* or algo*) ) OR AB=(semi supervised NEAR/2 (learning or model* or net* or algo*) )<br>14. TI=(deep hybrid NEAR/2 (learning or model* or net* or algo*) ) OR AB=(deep hybrid NEAR/2 (learning or model* or net* or algo*) )<br>15. TI=(bayes* NEAR/2 (learning or model* or net* or algo*) ) OR AB=(bayes* NEAR/2 (learning or model* or net* or algo*) )<br>16. TI=(cluster* NEAR/2 (learning or model* or net* or algo*) ) OR AB=(cluster* NEAR/2 (learning or model* or net* or algo*) )<br>17. TI=(gauss* NEAR/2 (learning or model* or net* or algo*) ) OR AB=(gauss* NEAR/2 (learning or model* or net* or algo*) )<br>18. TI=((feature NEAR/2 (engineer* or select* or extract* or learn*) ) or hyperparameter) OR AB=((feature NEAR/2 (engineer* or select* or extract* or learn*) ) or hyperparameter)<br>19. TI=(((neural or conv*) NEAR/0 (net* or learn* or model*) ) or CNN or convnet or RNN, or long short-term memory or lstm or gate* recurrent unit or gru or boltzmann machine or deep belief net* or spatial transformer net* or sum product network) OR AB=(((neural or conv*) NEAR/0 (net* or learn* or model*) ) or CNN or convnet or RNN, or long short-term memory or lstm or gate* recurrent unit or gru or boltzmann machine or deep belief net* or spatial transformer net* or sum product network) |

20. TI=(ensemble NEAR/0 (learning or model* or net* or algo*) ) OR AB=(ensemble NEAR/0 (learning or model* or net* or algo*) )
21. TI=(transfer NEAR/0 (learning or model* or net* or algo*) ) OR AB=(transfer NEAR/0 (learning or model* or net* or algo*) )
22. TI=(zero shot NEAR/0 (learning or model* or net* or algo*) ) OR AB=(zero shot NEAR/0 (learning or model* or net* or algo*) )
23. TI=(reinforcement NEAR/0 (learning or model* or net* or algo*) ) OR AB=(reinforcement NEAR/0 (learning or model* or net* or algo*) )
24. TI=(dictionary NEAR/0 (learning or model* or net* or algo*) ) OR AB=(dictionary NEAR/0 (learning or model* or net* or algo*) )
25. TI=((vector machine or SVM or ((classification or regression or probability or decision) NEAR/0 tree*) or random forest)) OR AB=((logistic regression or vector machine or SVM or ((classification or regression or probability or decision) NEAR/0 tree*) or random forest))
26. TI=((generative model* or autoencod* or aae or vae or cae or dae or sdae or gan or generative adversarial)) OR AB=((generative model* or autoencod* or aae or vae or cae or dae or sdae or gan or generative adversarial))
27. TI=(pca or principal component analysis or (k near/0 means) or (nearest near/0 neighbo$r) or knn or (fuzzy near/0 logi*) or isolation forest or hidden markov model or association rule* or feature bag* or score normali$ation) OR AB=(pca or principal component analysis or (k near/0 means) or (nearest near/0 neighbo$r) or knn or (fuzzy near/0 logi*) or isolation forest or hidden markov model or association rule* or feature bag* or score normali$ation)
28. #27 OR #26 OR #25 OR #24 OR #23 OR #22 OR #21 OR #20 OR #19 OR #18 OR #17 OR #16 OR #15 OR #14 OR #13 OR #12 OR #11 OR #10 OR #9
29. #28 AND #8

Cochrane library

| ID | Search |
|---|---|
| #1 | MeSH descriptor: [Magnetic Resonance Imaging] explode all trees |
| #2 | (mri or mr or (magnetic adj1 resonance)):ab,ti |
| #3 | MeSH descriptor: [Tomography, X-Ray Computed] explode all trees |
| #4 | (ct or (comput* adj1 tomograph*)):ab,ti |
| #5 | (cat adj3 (scan* or imag* or stud*)):ab,ti |
| #6 | neuroimag*:ti,ab |
| #7 | {OR #1-#6} |
| #8 | (brain or head or skull or cerebral or intracerebral or cerebrum or cranial or intracranial or cranium):ti,ab |
| #9 | (detect* or classif* or identif* or diagnos* or predict* or decision or decis* or decid*):ti,ab |
| #10 | {AND #7-#9} |
| #11 | MeSH descriptor: [Diagnosis, Computer-Assisted] explode all trees |
| #12 | MeSH descriptor: [Algorithms] explode all trees |
| #13 | MeSH descriptor: [Artificial Intelligence] explode all trees |
| #14 | MeSH descriptor: [Neural Networks, Computer] explode all trees |
| #15 | MeSH descriptor: [Machine Learning] explode all trees |
| #16 | ((artificial adj1 intelligence) or ((deep or machine) adj1 learning)) |
| #17 | algorithm*:ab,ti,kw or automat*:ab,ti or radiomic*:ab,ti,kw or (comput* adj3 (aid* or assist* or vision*)):ab,ti |
| #18 | ((supervised or unsupervised or (semi adj1 supervised) or deep hybrid or cluster* or bayes* or gauss*) adj3 (learning or model* or net* or algo*)) |
| #19 | ((feature adj3 (engineer* or select* or extract* or learn*)) or hyperparameter) |
| #20 | (((neural or conv*) adj1 (net* or learn* or model*)) or CNN or convnet or RNN, or long short-term memory or lstm or gate* recurrent unit or gru or boltzmann machine or deep belief net* or spatial transformer net* or sum product network) |
| #21 | ((ensemble or transfer or zero shot or reinforcement or dictionary) adj1 (learning or model* or net* or algo*)) |
| #22 | (vector machine or SVM or ((classification or regression or probability or decision) adj1 tree*) or random forest) |
| #23 | (generative model* or autoencod* or aae or vae or cae or dae or sdae or gan or generative adversarial) |
| #24 | (pca or principal component analysis or (k adj1 means) or (nearest adj1 neighbo?r) or knn or (fuzzy adj3 logi*) or isolation forest or hidden markov model or association rule* or feature bag* or score normali$ation) |
| #25 | {OR #11-#24} |
| #26 | #10 AND  #25 |

## SUPPLEMENTARY MATERIAL 2: FULL DATA EXTRACTION AND DATA ANALYSIS STRATEGY

### Inclusion criteria

We included studies where an AI model predicted the presence of an abnormality using a CT or MRI examination. To prevent major methodological bias related to generalisability, eligible studies included were those validating the AI model on a test dataset separated from the dataset used for development, either by time as a minimum acceptable inclusion standard (temporal validation), or by location (geographical validation; defined as external validation in this review).

In routine clinical practice, imaging patient cohorts contain normal brains and a range of pathological conditions. For AI studies detecting a single pathological target condition such as intracranial hemorrhage, a representative study would require that the AI model was validated not only on cohorts that contained the target condition, but also a reasonably representative range of non-target conditions that might be obtained during routine scanning. As a minimum acceptable inclusion standard, eligible studies were required to have test datasets that contained normal scans, scans containing the target, and one or more non-target conditions.

### Exclusion criteria

The motivation for the study was to review abnormality detection in first-line, clinical neuroimaging.

Angiography or perfusion studies, for example, were excluded on the grounds that they would typically be used as second-line investigations triggered by a high clinico-radiological suspicion for a particular pathology (e.g., CT angiography in patients with clinico-radiological features of stroke caused by a large vessel occlusion). For the same rationale, other advanced MR techniques (e.g., MR spectroscopy) were excluded.

Target conditions with a structural correlate often seen at the individual level (e.g. Alzheimer's disease) were included. In contrast, those conditions where structural differences have been shown only in group-wise comparison studies but are not apparent to the radiologist at the individual level, were excluded (e.g., a range of psychiatric conditions where voxel-based morphometry has shown structural differences when compared to healthy controls).

All studies using other forms of internal validation not adhering to the minimum acceptable inclusion standard of temporal validation, were excluded if there was no external validation. An example of this scenario is studies using cross-validation alone. Studies that only reported the accuracy of the AI model to make voxel-wise (e.g., segmentation studies) or slice-wise predictions but did not subsequently report at the patient level were excluded (unless patient level accuracy could be calculated from the published study data).

Studies testing exclusively on pediatric populations were excluded. Studies not published in a peer-reviewed journal or without an English language translation were excluded[1].

## Study selection

S.A. and D.W. (radiologist and data scientist, both with 3 years neuroimaging research experience), independently reviewed the titles and abstracts of all retrieved records using the inclusion criteria. Full texts were then assessed for eligibility with any final arbitration through a third reviewer (T.C.B., neuroradiologist, 16 years neuroimaging research experience).

## Data extraction and quality assessment

Study quality was assessed independently (S.A. and D.W.) with any final arbitration through a third assessor (T.C.B.). We used the QUality Assessment of Diagnostic Accuracy Studies-2 (QUADAS-2) tool[2], tailored to the review question incorporating items from the Checklist for Artificial Intelligence in Medical Imaging (CLAIM)[3]; modified signalling questions are presented in Supplementary Material 3.

## Data analysis

The unit of analysis was the patient undergoing a CT or MRI examination. The primary outcome was diagnostic test accuracy. Secondary outcomes assessed whether the AI model had been applied in a clinical pathway, and if so, the associated performance metrics.

To determine the primary outcome measures, where published, the 2 x 2 contingency tables and the principal diagnostic accuracy measures of sensitivity (recall) and specificity were extracted for test datasets. The area under the receiver operating characteristic curve (ROC-AUC) values, and positive predictive values (PPV or precision) were also extracted where published. Where 2 x 2 contingency tables were not provided, the tables were populated based on the published study data; the calculations are outlined in Supplementary Material 4.

The PPV is important for abnormality detection (Supplementary Material 4) and is more informative than specificity for imbalanced datasets where the prevalence of the target condition is small[4]. To achieve this, the PPV is adjusted for direct comparison of AI models. There were sufficient studies in one subgroup (intracranial hemorrhage detection on CT scans) for the calculation of prevalence-adjusted PPV adjustment. Here, we chose a prevalence of 10% based on recent evidence of routine clinical practice in the UK[5]; the calculation is outlined in Supplementary Material 4[6].

### *Meta-analysis*

Meta-analysis was performed when four or more studies evaluated a given target condition within a specific modality[7]. Studies investigating the detection of intracranial hemorrhage on CT scans were the only subgroup included for meta-analysis.

A bivariate random-effects model was used for meta-analysis, taking into account the within and between study variance, and the correlation between sensitivity and specificity across studies[8]. Sensitivities and specificities were presented for each study using forest plots, and pooled estimates for both measures were calculated. To investigate the impact of variables of interest contributing to heterogeneity, meta-regression was performed with the variable of interest as a covariate for the bivariate model. Using the existing model parameters, the absolute differences in pooled sensitivity and specificity between subgroups of interest were computed.

Parameters of the model also allowed for the estimation of the summary ROC (SROC) curve and the summary ROC-AUC (SROC-AUC). Using a resampling approach[9], the model estimates were also used to derive the pooled measures of balanced accuracy as well as the positive and negative likelihood ratios and the diagnostic odds ratio.

The meta-analysis was conducted by a statistician, M.G., with 15 years of relevant experience. All the statistical analyses were performed in R (v 3.6.1). The R package mada (v 0.5.10) was used for the bivariate model[10]. Since some of the 2 x 2 contingency table input cell values derived from the individual studies contained zeros, we applied a continuity correction (0.5).

## SUPPLEMENTARY MATERIAL 3: MODIFIED QUADAS-2 SIGNALLING QUESTIONS

| Question | Points specific to this review |
|---|---|
| **PARTICIPANT SELECTION – A. RISK OF BIAS** | |
| Was a consecutive or random sample of patients enrolled in the test set? | Yes: datasets of unenriched, consecutive patients OR randomly selected patients<br>Unclear: Not stated<br>No: Other |
| Did the study avoid inappropriate exclusions and does the study likely have a range of pathology matching a clinical cohort? | Yes: Clinical cohorts with no inappropriate exclusions<br>Unclear: Not stated<br>No: 3 or fewer categories of pathology |
| Was the test dataset geographically distinct from the training dataset | Yes: Geographically distinct<br>Unclear: Not stated<br>No: Temporally distinct only OR Internal validation only: e.g. hold-out, cross-validation |
| **PARTICIPANT SELECTION - B. CONCERNS REGARDING APPLICABILITY** | |
| Is there a concern that the included patients do not match the review question? | High concern if any of:<br>• A consecutive or random sample of patients weren't used<br>• An enriched sample not matching the prevalence of disease<br>• 3 or fewer categories of pathology<br>• Any pathology or subtype of pathology excluded |
| **INDEX TESTS – A. RISK OF BIAS** | |
| Were patients present in the test dataset not used for development? | Yes: The same patients are not in both test and training datasets<br>No: Evidence of data leakage |
| If a threshold was used, was it pre-specified? | Yes: Threshold set previously e.g. in a commercially available AI system, or threshold or operating point set during training<br>Unclear: Not stated<br>No: Accuracy metrics provided directly on the test dataset, without pre-specifying operating point during model development |
| **INDEX TESTS - B. CONCERNS REGARDING APPLICABILITY** | |
| Are there concerns that the index test, its conduct, or interpretation differ from the review question? | High concern if any of:<br>Analytical validation only (in laboratory conditions/analytical validation)<br>Low concern if:<br>AI deployed in clinical practice (clinical validation) |
| **REFERENCE STANDARD – A. RISK OF BIAS** | |
| Is the reference standard likely to correctly classify the target condition? | Yes if any of:<br>• 2 or more radiologists independently reviewed the images in a study, for each |
| | study in the test set. If labels were derived from existing radiology reports by 1 radiologist, a different radiologist must have reviewed the images.<br>• Subsequent information available from different modality or histopathology e.g. DWI-MR providing the labels for ischemic stroke in CT<br>No: Fewer than 2 radiologists reviewed images for each study (e.g. 1 radiologist extracted labels from existing reports) |
| Were the reference standard results interpreted without knowledge of the results of the index test? | Yes: AI output was not known to individual(s) creating the reference standard<br>No: Reference standard informed by AI output |
| **REFERENCE STANDARD - B. CONCERNS REGARDING APPLICABILITY** | |
| Are there concerns that the target condition as defined by the reference standard does not match the review question? | High concern if any of:<br>• Labels in the test dataset were not generated by a radiologist (e.g. not a natural language processing model)<br>• Findings expected in healthy ageing (e.g. cerebral atrophy and small vessel disease commensurate for age) are considered abnormal by the reference standard |
| **FLOW AND TIMING – A. RISK OF BIAS** | |
| Were all patients included in the analysis? | Yes: Patients in final analysis the same as in the initial sample.<br>No: Differing number of patients between sampling and results without adequate explanation. |

## SUPPLEMENTARY MATERIAL 4: CALCULATIONS FOR PREVALENCE ADJUSTED PPV, CONFUSION MATRICES FROM PROVIDED ACCURACY METRICS

The positive predictive value (PPV) is preferred to specificity in abnormality detection, as it is a measure of how well the target condition is correctly identified from all scans predicted to be positive by an AI model (true positives and false positives). Comparatively, specificity focuses on how well the normal class is identified from all scans that are negative for the target condition (true negatives and false positives). In extremely imbalanced datasets where the normal class has a far greater prevalence than the target condition, an AI model that classifies all data as 'normal' would have reasonable specificity, but poor PPV. PPV, however, is dependent on prevalence, therefore a prevalence-adjusted PPV needs to be calculated to compare AI models detecting the same target condition.

1. Calculating prevalence-adjusted positive predictive value (PPV) from sensitivity and specificity[11]:

$$PPV = \frac{sensitivity \cdot prevalence}{(1 - specificity)(1 - prevalence) + (sensitivity \cdot prevalence)}$$

If the prevalence of the target condition is consistently 10% across the compared studies, the prevalence-adjusted PPV can be calculated as follows:

$$Prevalence\ adjusted\ PPV = \frac{sensitivity \cdot 0.1}{(1 - specificity)(1 - 0.1) + (sensitivity \cdot 0.1)}$$

2. Calculating the number of true positives (TP), false negatives (FN), false positives (FP), true negatives (TN) from sensitivity, specificity, the total number of scans (total) and the number of scans with the target condition in the test dataset (P):

$$N = total - P$$

$$TP = sensitivity \cdot P$$

$$TN = specificity \cdot N$$

$$FN = P - TP$$

$$FP = N - TN$$

3. Calculating the number of true positives (TP), false negatives (FN), false positives (FP), true negatives (TN) from sensitivity, specificity, the PPV the total number of scans in the test dataset (total):

$$TP = \frac{total}{\left(\frac{1}{PPV} - 1\right) \cdot \frac{specificity}{1 - specificity} + \frac{1}{sensitivity} + \frac{1}{PPV} - 1}$$

$$FN = TP \cdot \left(\frac{1}{sensitivity - 1}\right)$$

$$FP = \frac{TP}{PPV} - TP$$

$$TN = TP \cdot \left(\frac{1}{PPV} - 1\right) \cdot \frac{specificity}{1 - specificity}$$

## SUPPLEMENTARY MATERIAL 5: SUMMARY OF CHARACTERISTICS FOR EACH STUDY.

| Study (author, year) | Modality, Target pathology | Index test | Training set | Test set | Reference standard (Ground truth for test set) |
|---|---|---|---|---|---|
| Arbabshirani (2018) | CT, ICH | CNN<br>Labels: Examination-level<br>Output: Binary prediction of ICH (present/not present) for each examination | 46,583 CT head examinations from 31,256 patients, from 2007-2017, from 1 institution (USA), 4 scanner vendors, randomly split into training (24,882) and internal hold out test sets (6,374). Training set labels extracted from clinical reports. | 347 non-urgent inpatient and outpatient CT head studies during a 3 month implementation in clinical practice in 2017. | 25% of radiologist reports were converted to labels by research assistants under supervision of one neuroradiologist (>11,600), 75% using an NLP algorithm. |
| Buls (2021) | CT, ICH | Aidoc v1.3, a proprietary CNN<br>Labels: examination-level, bounding boxes, segmentation<br>Output: Binary prediction of ICH (present/not present) for each examination, key images for review | Approximately 50,000 CT head examinations from 9 different sites, 17 scanner models. Ground truth labels varied depending on hemorrhage type and size, including study-level binary labels, slice-level bounding boxes and voxel-level segmentation. | A subset of 388 out of 500 CT head studies consecutively acquired from Sep - Oct 2019 at 1 institution in Brussels (Belgium), 4 scanners, 3 scanner vendors, with patients <18 years of age excluded. 112 studies were not done because the AI model could not process them in real-time, although the cause of failure was not investigated. | Consensus opinion for each study derived by three neuroradiologists with between 5 - 15 years of experience, with access to previous studies, clinical history and clinical reports. |
| Chang (2018) | CT, ICH | CNN, modified mask R-CNN architecture<br>Labels: examination-level, segmentation<br>Outputs: Binary prediction of ICH (present/not present) for each examination, segmentation and volume estimation of ICH | 10,159 CT head examinations, from 1 institution from January to August 2017. Examination-level labels created manually. | All 682 emergency department CT head examinations, from the same institution, consecutively acquired in February 2018. | ICH positive cases were identified from clinical reports and confirmed visually by a radiologist. Segmentation masks were generated semi-automatically by a radiologist. |
| Chilamkurthy (2018) | CT, ICH<br>CT, mass effect | Qure.ai proprietary CNN, modified ResNet18 architecture<br>Labels: slice-level<br>Output: Confidence score for each slice in examination, separate random forest machine learning model used to convert confidence for each slice into binary prediction of ICH for each examination | 313,318 CT head examinations from 2011-2017, collected from 20 institutions (India), 3 scanner vendors, 12 scanner models, randomly split into training (290,055) and internal hold out test sets (21,095). Slice-level labels were manually created for ICH (4304 scans, 165809 slices), midline shift and mass effect (699 scans, 26135 slices). | "CQ500 dataset": Enriched dataset of 491 studies from 2012-2018, from 6 institutions in New Dehli (India), 2 scanner vendors, 6 scanner models, with postoperative patients and patients <7 years of age excluded. 214 scans consecutively collected in Nov 2017, 277 scans selected for ICH present in the report using an NLP tool. | Three radiologists independently recorded the presence of ICH, midline shift and fracture for all 491 studies, with majority vote deciding the labels for each study. |
| Chilamkurthy (2018) | CT, Skull fracture | Qure.ai proprietary CNN, modified DeepLab architecture<br>Labels: Bounding-box annotations per slice<br>Output: Confidence score for each slice in examination, separate random forest model | | | |
| | | used to convert into binary prediction of fracture for each examination | Bounding-box labels were manually created for fracture (1119 scans, 9938 slices) | | |
| Finck (2021) | CT, any pathology | "Weakly supervised machine learning": normative learning by registering normal brains to a shared atlas and determining per-voxel confidence-intervals<br>Labels: N/A - only trained on normal brains<br>Output: Anomaly heat map: voxels where value was outside the CIs, Anomaly score: ratio of outlier voxels to entire brain ranging from 0 to 1, Prediction of any pathology for each examination into three classes: normal, uncertain, abnormal | 191 normal CT head examinations, from 2018 - Feb 2020, 1 institution (Germany), 1 scanner model for training. 31 pathological and 30 normal CT head examinations, 2018 - 2019, for validation. Examination-level labels assigned. | 248 consecutive CT head studies, Mar 2020, 1 institution, 1 scanner model, after excluding follow up examinations (170) and patients with metal implants (56). | Two neuroradiologists reviewed the images and report for each study to extract an examination-level label |
| Ginat (2020) | CT, ICH | Aidoc v1.3 (see Buls 2018 above) | Aidoc v1.3 (see Buls 2018 above) | 2,011 consecutively acquired urgent CT head studies, Jan - Feb 2019, from 1 tertiary hospital, 9 scanners | One neuroradiologist reviewed the images and clinical report for each study |
| Ginat (2021) | CT, ICH | Aidoc (see Buls 2018 above) | Aidoc (see Buls 2018 above) | 8,723 consecutively acquired CT head studies, May 2020 - Feb 2021, from 1 tertiary hospital, 9 scanners | 'The final radiologist report defined the ground truth' – implied that same radiologist that reported extracted the binary examination-level label |
| Kuo (2019) | CT, ICH | CNN, 'PatchFCN' (modified ResNet-38 architecture)<br>Labels: Lesion segmentations<br>Output: Lesion segmentations, prediction of ICH (present/not present) for each examination | 4,396 CT head studies from 2010-2017 of which 1,131 were positive for ICH, collected from affiliated hospitals of UCSF (USA), 2 scanner vendors, trained using randomly split 4-fold cross validation. Voxel-wise ICH segmentations, manually performed, used as labels for 1,131 positive scans. Studies required 'skull-stripping' as a pre-processing step. | 200 CT head studies performed at the same hospitals, November - December 2017. Excluding previous neurosurgery, from all CT head scans performed, 150 were randomly chosen if no previous studies were performed in the same visit, 50 were randomly chosen if more than one CT head was performed on the same visit. Skull-stripping failed on 1 study, which was replaced by another from the same time period. | Consensus opinion of two neuroradiologists for each examination |
| Monteiro (2020) | CT, ICH | CNN, DeepMedic architecture<br>Labels: Lesion segmentations<br>Output: Lesion segmentations, binary prediction of ICH (present/not present) for | 839 CT head studies from 512 patients, from 2014 - 2017, 60 institutions across Europe, subset of CENTER-TBI study, randomly split into training (184) and | CQ500 (see Chilamkurthy, 2018) | CQ500 (see Chilamkurthy, 2018) |
| | | each examination created by only considering segmentations >1ml as ICH | internal hold out test sets (655). Segmentation labels semi-automatically generated by manually correcting outputs of an earlier version of the AI model. | | |
| McLouth (2021) | CT, ICH | Avicenna.ai, CINA v1.0: proprietary AI model<br>Labels: not disclosed<br>Output: Binary prediction of acute, hyperdense ICH (present/not present) for each examination | 8,994 CT head studies, multiple institutions through vRAD, a teleradiology service (USA), 2014-2018. The AI model 'was only trained to identify acute blood based off of hyperdense components', 'chronic hemorrhages cannot be identified by the AI model unless they contain more acute hyperdense components' | Patients were pre-selected for suspected ICH by search terms in the clinical indication e.g. 'hemorrhage', 'NCCT', 'head'. 395 CT head studies from vRAD (USA), 2019, 419 studies from University of California, Irvine (USA), 2017-2019, although sampling method was unclear. 4 scanner manufacturers. 10 studies excluded for insufficient image quality or contrast-enhanced. | Consensus of two neuroradiologists, with a third for arbitration. |
| Prevedello (2017) | CT, ICH, mass effect, hydrocephalus ('algorithm 1') | CNN, modified GoogLeNet architecture<br>Labels: Examination-level<br>Output: Binary prediction of pathology (present/not present) for each examination | 246 CT head studies, 1 institution (USA), >1 scanner model, 146 positive findings from a consecutive sample and 100 healthy controls from a random sample, from which 2583 2-D images with the representative pathology were split randomly into training (80%) and internal hold out test sets. Labels were extracted manually from clinical report. | 130 CT head studies from a consecutive sample of 226 studies in 2015, after exclusion of postoperative patients and degraded images. | Manually extracted from clinical report by one radiologist |
| Prevedello (2017) | CT, acute infarct ('algorithm 2') | | 71 CT head studies, from the training sample of algorithm 1, that did not have hemorrhage, mass effect or hydrocephalus, 46 with pathology and 25 healthy controls, with 2-D images split randomly into training (80%) and internal hold out test sets. Labels were extracted manually from clinical report. | 49 CT head studies from the same sample, that in practice would be the negative cases from algorithm 1. It was unexplained why only 19 normal cases were analysed by algorithm 2 when 47 of the normal cases were marked negative by algorithm 1. | |
| Salehinejad (2021) | CT, ICH | CNN, ensemble model of modified ResNeXt-50 and ResNeXt-101 architectures. | 21,784 CT head studies, 1999-2018, 3 institutions (Brazil, USA), multiple scanner models. Slice level labels for 674,258 slices | 5,965 consecutively acquired emergency CT head studies, Jan 2019 - Dec 2019, from 1 tertiary | Manually extracted from neuroradiologist report by research assistant - sample of 600 reviewed by radiologist |
| | | Labels: slice-level<br>Output: Binary prediction of ICH (present/not present) for each examination | created by over 60 neuroradiologists for RSNA 2019 “Brain CT Hemorrhage Challenge”. | hospital, 3 models from same manufacturer. | with 100% examination-level labels correct, 98.1% correct for ICH subtype |
| Wang (2021) | CT, ICH | Ensemble model of CNN and two recurrent neural networks. Winner of the 2019-RSNA “Brain CT Hemorrhage Challenge”.<br>Labels: slice-level<br>Output: Binary prediction of ICH (present/not present) for each slice and examination | 19,530 CT head studies, 1999-2018, 3 institutions (Brazil, USA), multiple scanner models. Slice level labels for 674,258 slices created by over 60 neuroradiologists for RSNA 2019 “Brain CT Hemorrhage Challenge” | CQ500 (see Chilamkurthy, 2018) | CQ500 (see Chilamkurthy, 2018) |
| Voter (2021) | CT, ICH | Aidoc (see Buls, 2018) | Aidoc (see Buls, 2018) | 3,605 consecutively acquired CT head studies, July 2019 - Dec 2019, from 1 tertiary hospital, 7 scanners from same manufacturer | One neuroradiologist reviewed the clinical report and the AI prediction to set the ground truth. |
| Gauriau (2021) | MR, any pathology | CNN<br>Labels: Examination-level<br>Output: Binary prediction of pathology (present/not present) for each examination | 2,741 axial MR FLAIR examinations, consisting of 1,987 adults collected from 2007-2017 from 1 institution and 5,808 female patients only, including children, collected in 2017 from multiple sites at another institution. These datasets were originally collected for other studies. | 1,489 consecutive MR axial FLAIR examinations collected in 2019, children included, multiple sites but some data came from the same institution as training data, 4 scanner vendors, 14 scanner models | Manually extracted from reports, and images independently reviewed by a radiologist from another institution. Discrepancies would be settled by consensus. |
| Nael (2021) | MR, any pathology | CNN, modified U-net architecture<br>Input: MR FLAIR, ADC, DWI<br>Labels: Examination-level<br>Output: Binary prediction of pathology (present/not present) for each examination | 12,143 MR examinations of adult patients, after exclusion for 'non-definitive labels' (86) and without mandatory sequences (1,851), randomly split into training (9,845), internal validation (1,248) and internal hold-out test set (1,050). 2 scanner vendors, 19 scanner models | 1,072 MR examinations of adult patients, after exclusion for lack of mandatory sequences (659), 2 external institutions, 3 scanner vendors | One neuroradiologist extracted examination-level labels based on the clinical report |
| Nael (2021) | MR, ICH | | | | |
| Nael (2021) | MR, acute infarct | | | | |

CT = computed tomography, NCCT = non-contrast computed tomography, MR = magnetic resonance, FLAIR = fluid-attenuated inversion recovery, DWI = diffusion weighted imaging, ADC = apparent diffusion coefficient. FLAIR, DWI and ADC are commonly used MR sequences. ICH = intracranial hemorrhage, CNN = convolutional neural network, NLP = natural language processing, vRAD = Virtual Radiologic (commercial teleradiology solution), RSNA = Radiological Society of North America. Aidoc, Qure.ai and Avicenna.ai are commercial vendors of AI products. Aidoc v1.0, Aidoc v1.3 and CINA v1.0 are commercial AI solutions. CQ500 is a publicly available test dataset of CT head examinations from 491 patients. mask R-CNN, PatchFCN, GoogLeNet, ResNet18, ResNet38, ResNeXt-50, ResNeXt-101, U-net, DeepLab and DeepMedic are CNN architectures, published in academic literature.

| Study (author, year) | Modality | Target pathology | Index test | Training labels | Model outputs |
|---|---|---|---|---|---|
| Arbabshirani (2018) | CT | ICH | CNN | Examination-level binary presence of abnormality (present/not present) | Binary prediction of ICH (present/not present) for each examination |
| Buls (2021) | CT | ICH | Aidoc v1.3, a proprietary CNN | Combination of examination-level binary labels, bounding boxes and segmentations | Binary prediction of ICH (present/not present) for each examination, key images for review |
| Chang (2018) | CT | ICH | CNN, modified mask R-CNN architecture | Manual segmentations for each examination | Binary prediction of ICH (present/not present) for each examination, segmentations, and volume estimation of ICH |
| Chilamkurthy (2018) | CT | ICH | Qure.ai proprietary CNN, modified ResNet18 architecture | Slice-level binary presence of abnormality (present/not present) | Binary prediction of ICH (present/not present) for each examination |
| | CT | Mass effect | | | |
| Chilamkurthy (2018) | CT | Skull fracture | Qure.ai proprietary CNN, modified DeepLab architecture | Bounding-box annotations per slice | |
| Finck (2021) | CT | Any pathology | “Weakly supervised machine learning”: normative learning by registering normal brains to a shared atlas and determining per-voxel confidence-intervals | Not directly trained on labels, although training was conducted on brains that were known to be normal | Prediction of any pathology for each examination into three classes: normal, uncertain, abnormal.<br>Anomaly score: ratio of outlier voxels to entire brain ranging from 0 to 1<br>Anomaly heat map: voxels where value was outside the CIs, |
| Ginat (2020) | CT | ICH | Aidoc (see Buls 2018 above) | | |
| Ginat (2021) | CT | ICH | | | |
| Kuo (2019) | CT | ICH | CNN, 'PatchFCN' (modified ResNet-38 architecture) | Manual segmentations for each examination | Binary prediction of ICH (present/not present) and lesion segmentations for each examination |
| Monteiro (2020) | CT | ICH | CNN, DeepMedic architecture | Semi-automatically created segmentations for each examination | Binary prediction of ICH (present/not present) and lesion segmentations for each examination (output segmentations >1ml were considered as ICH present) |
| McLouth (2021) | CT | ICH | Avicenna.ai, CINA v1.0: proprietary AI model | Not disclosed | Binary prediction of acute, hyperdense ICH (present/not present) for each examination |
| Prevedello (2017) | CT | ICH, mass effect, hydrocephalus (‘algorithm 1’) | CNN, modified GoogLeNet architecture | Examination-level presence of abnormality (present/not present) | Binary prediction of pathology (present/not present) for each examination |
| Prevedello (2017) | CT | Acute infarct (‘algorithm 2’) | | | |
| Salehinejad (2021) | CT | ICH | CNN, ensemble model of modified ResNeXt-50 and ResNeXt-101 architectures. | Slice-level binary presence of abnormality (present/not present). | Binary prediction of ICH (present/not present) for each examination |
| Wang (2021) | CT | ICH | Ensemble model of CNN and two recurrent neural networks. Winner of the 2019-RSNA “Brain CT Hemorrhage Challenge”. | Slice-level binary presence of abnormality (present/not present) | Binary prediction of ICH (present/not present) for each slice and examination |
| Voter (2021) | CT | ICH | Aidoc (see Buls, 2018) | | |
| Gauriau (2021) | MR | Any pathology | CNN | Examination-level binary presence of abnormality (present/not present) | Binary prediction of pathology (present/not present) for each examination |
| Nael (2021) | MR (FLAIR, ADC, DWI) | Any pathology | CNN, modified U-net architecture | Examination-level binary presence of abnormality (present/not present) | Binary prediction of pathology (present/not present) for each examination |
| Nael (2021) | MR (FLAIR, ADC, DWI) | ICH | | | |
| Nael (2021) | MR (FLAIR, ADC, DWI) | Acute infarct | | | |

CT = computed tomography, MR = magnetic resonance, FLAIR = fluid-attenuated inversion recovery, DWI = diffusion weighted imaging, ADC = apparent diffusion coefficient. FLAIR, DWI and ADC are commonly used MR sequences. ICH = intracranial hemorrhage, CNN = convolutional neural network, RSNA = Radiological Society of North America. Aidoc, Qure.ai and Avicenna.ai are commercial vendors of AI products. Aidoc v1.0, Aidoc v1.3 and CINA v1.0 are commercial AI solutions. Mask R-CNN, PatchFCN, GoogLeNet, ResNet18, ResNet38, ResNeXt-50, ResNeXt-101, U-net, DeepLab and DeepMedic are CNN architectures, published in academic literature. Training labels are inputs to machine learning models

## SUPPLEMENTARY MATERIAL 6: FURTHER DETAILS FOR THE ASSESSMENT OF RISK OF BIAS

### Patient Selection

Five studies (5/16, 31%) were considered overall to be at low risk of bias for patient selection [12–16].

Seven studies (7/16, 44%) that used temporal validation alone without external validation, were considered to have a high risk of bias for patient selection as there is limited assessment of generalisability [17–23], compared to 9/16 (56%) studies where AI models were externally validated on test data from other institutions [12–16,24–27].

Ten studies (10/16, 63%) with consecutive sampling of cases for test datasets [12–18,20,21,23] were considered to have a low risk of bias for this attribute. The sampling method was unclear in one study (1/16, 6%) [27]. A high risk of bias was found for the remaining four studies (4/16, 25%) where target condition cases were added to enrich test datasets, which were therefore unrepresentative of the prevalence of abnormalities encountered in clinical practice [19,24–26]. Similarly, there was a high risk of bias in one intracranial hemorrhage detection study (1/16, 6%) which excluded patients in the test dataset if the clinical request for the scan did not state the word "hemorrhage" [22]; we considered this an inappropriate exclusion.

### Index test

Researchers can choose different "operating points", by adjusting the threshold at which the continuous output of their AI model is converted into a binary decision for abnormality detection. This allows for the same model to be tuned to favour, for example, either sensitivity or specificity. Three studies (3/16, 19%) were considered at high risk of bias for selecting the operating point after testing as it allows researchers to present optimised accuracy metrics, and fails to convey the reliability of the AI model at the threshold chosen during model development, prior to testing [19,24,25]. It was unclear whether an operating point has been selected before testing in 3/16 (19%) studies [20,23,26].

There were concerns regarding applicability in eight studies (8/16, 50%) as they assessed AI model performance in laboratory conditions ("analytical validation"[28]) only [16,19,21,22,24–27]. In contrast, four studies (4/16, 25%) placed the AI model within the clinical pathway ("clinical validation") [12–14,17], which more closely resembles a "real world" environment and therefore the intended applicability. In the remaining four studies (4/16, 25%), it was unclear whether the AI model was validated clinically or in laboratory conditions [15,18,20,23].

### Reference Standard

The reference standard in all studies was radiologist assessment. Studies that used fewer than two radiologists to assess the images of a scan for their reference standard were considered at high risk of bias, as individual radiologists do not have perfect accuracy. For example, one study reported that the agreement between two neuroradiologists to label MRI brain reports as normal or abnormal was 94.9% [29]. Five studies (5/16, 31%) were therefore considered to have high risk of bias as only the clinical report was reviewed (in these cases, the report was assessed by either a single radiologist [13,20,27], research assistant [15,17] or an automated natural language processing algorithm [17]). 10 (10/16, 56%) studies were at a low risk of bias for this attribute as at least two radiologists had reviewed the images. In 2/16 (13%) studies, a single radiologist assessed both the images and the clinical report written by a different radiologist [12,23]. Of the remaining studies with a low risk of bias for this attribute, 3/16 (19%) used the "majority vote" of three independent radiologists as the reference standard [24–26] and 5/16 (31%) used a consensus of multiple radiologists (ranging from 2-4) [14,18,19,21,22]. A high risk of bias was additionally found for one study (1/16, 6%) as the reference standard was informed by the output of the AI model (the index test) [16]; the study was excluded from the meta-analysis due to this fundamental methodological flaw.

Of the three studies that aimed to detect any pathology, 2/16 (13%) did not consider healthy ageing leading to age-appropriate brain volume loss as normal [30] which raised concerns for applicability. We noted that there was a high prevalence of pathology in both test samples, 64% [21] and 81% [27] respectively, which may have been influenced by this decision. It was unclear how healthy aging was considered in one study (1/16, 6%); we note 72% of their test sample was considered abnormal [18].

## Flow and timing

There was a high risk of bias for flow and timing in two studies (2/16, 13%) where there was a discrepancy between the cases in the final analysis and the initial sample (e.g., due to AI model processing failures some patients were excluded from the final analysis) [14,19]. There were discrepancies in two studies (2/16, 13%) between the published data and the published contingency tables without adequate explanation, which were also considered as having a high risk of bias [17,20]. One study had two test datasets. One of the test datasets consisted of only 29% (140/491) of the CQ500 dataset without further explanation, therefore we excluded this dataset from analysis; we included the other test dataset which was from their own institution and was temporally distinct from their training data [18].

## SUPPLEMENTARY MATERIAL 7: DIRECT COMPARISON TO RADIOLOGISTS

The CQ500 dataset is a publicly available dataset of CT heads obtained from a representative clinical cohort and then enriched through the addition of intracranial hemorrhage cases, resulting in 205/491 cases of intracranial hemorrhage[24]. The reference standard for the CQ500 dataset is the "majority vote" from three radiologists independently reviewing each examination under laboratory conditions. When an AI model is undergoing validation and is being compared to the reference standard, there is also the opportunity for the AI model to be compared to the three individual radiologists, whose performance accuracy is also open source. Three studies (3/16, 19%) used the CQ500 dataset for AI model testing during external validation[24–26]. A statistical comparison was available for one study which showed equivalent sensitivity ($p = 0.86$) between radiologists and the AI model but with poorer specificity ($p < 0.001$)[24].

In another study with a small test dataset of 200 cases containing 25 cases of intracranial hemorrhage, the performance of a temporally validated AI model[19] was compared with four radiologists, who independently reviewed each examination under laboratory conditions. Separately, the consensus of two different neuroradiologists was used as the reference standard. No statistical comparison was available.

The performance of AI models and the individual radiologists from these four studies are highlighted in Figure 5.

England's National Institute for Health and Care Excellence (NICE) aligns professional standards through clinical guidelines and offers evidence-based guidance globally[31]. The findings of our review, unlike those of NICE, did not show that AI products are as effective at detecting intracranial hemorrhages as neuroradiologists[32]. Direct comparison to radiologists was available in four intracranial hemorrhage detection studies. Only one externally validated AI model appeared to show comparable performance to individual radiologists, albeit under laboratory conditions[26]. Another study showed a statistically comparable sensitivity to pooled radiologists but with inferior specificity[24]. It should be noted however that evidence from mammography screening suggests that radiologist performance in the laboratory is significantly worse than in a clinical setting (the 'laboratory effect') which is thought to be due to the lower stakes involved[33].

## SUPPLEMENTARY MATERIAL 8: COMMERCIAL SOLUTIONS

Many commercial AI solutions exist for brain imaging. In CT imaging, for example, 11 vendors provide commercially available (CE-marked) AI products for intracranial hemorrhage detection, but only four products have had their accuracy metrics published in peer-reviewed journals[34,35]. This lack of transparency is indicative of a wider problem in commercially available AI in medical imaging – a recent review found that of the available CE-marked AI products, only 36% had peer-reviewed evidence and 18% demonstrated clinical impact in the broadest sense[35].

Of the commercially available solutions included in this review, Aidoc demonstrated consistent accuracy when clinically validated in three independent patient cohorts in Belgium and the US[12–14]. By comparison, studies validating AI from Qure.ai and Avicenna.ai provided their own external test datasets, which is a potential source of bias. Another potential source of bias is that the model trained by Qure.ai used a training dataset that was exclusively sampled within India – further validation in other countries would be required to demonstrate generalisability to other countries where it is marketed[24]. The model from Avicenna.ai was only trained on a subset of acute intracranial hemorrhage, and was only tested on a subset of CT head examinations that specifically stated the word “hemorrhage” in the clinical request[22]. Clinically, non-acute intracranial hemorrhage, often called “haematoma”, can share many of the same management implications as acute intracranial hemorrhage. Furthermore, confining the detection to a population with a high clinical suspicion of the target condition, also limits applicability in routine clinical practice as this is not representative.

# SUPPLEMENTARY FIGURES

Supplementary Figure 1: PRISMA flow diagram

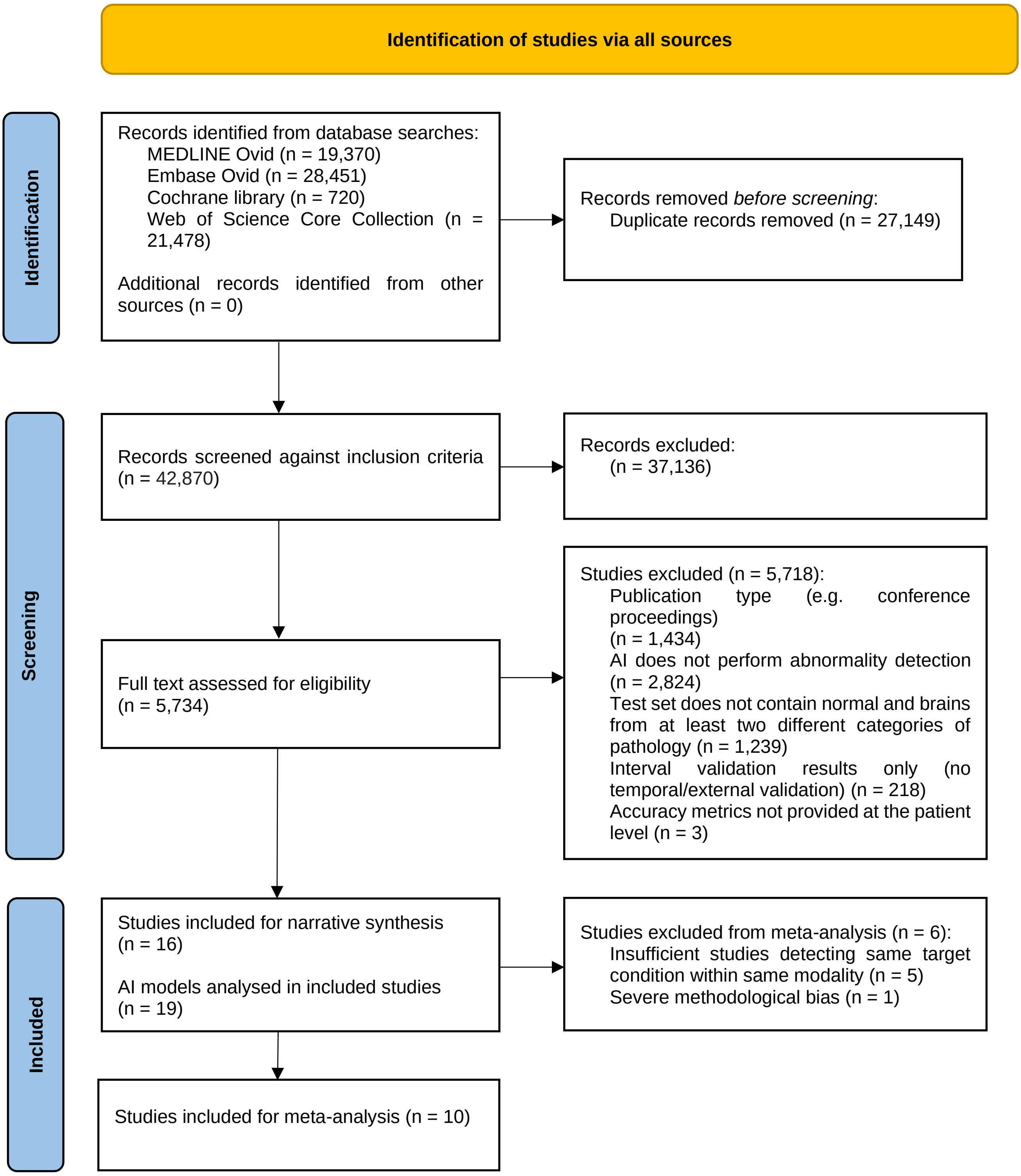

Supplementary Figure 2: Diagnostic test accuracy of AI models in CT imaging in receiver operating characteristic (ROC) space, compared with individual radiologists. An ideal classifier would be at the top-left corner at (0, 1). The size of each marker is proportional to the size of the test dataset. In studies where there was more than one operating point, we chose the operating point with the highest sensitivity. ICH = intracranial haemorrhage. CQ500 = CQ500 external test dataset. Qure.ai, Aidoc and Avicenna.ai are commercial vendors for AI products.

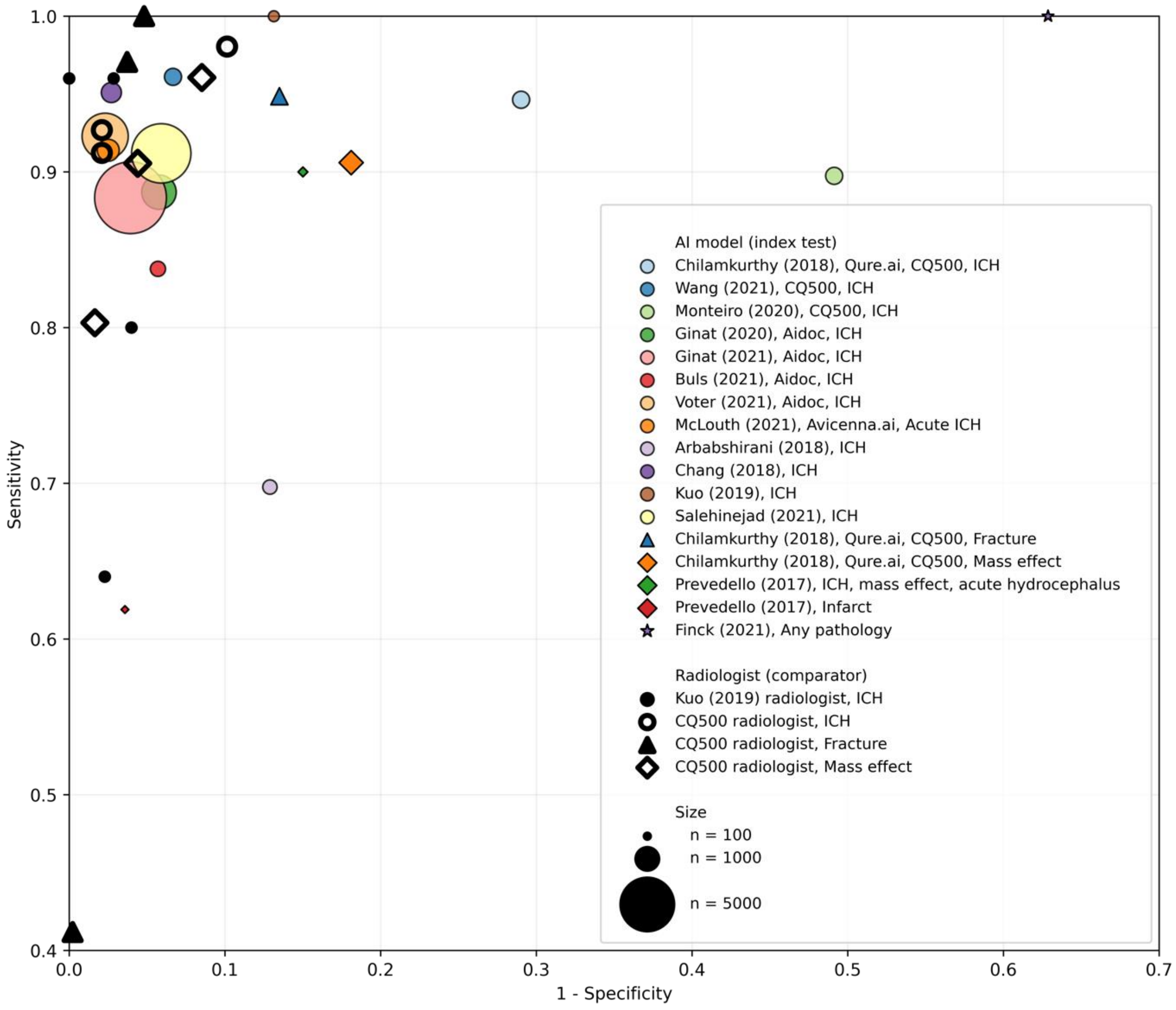

Supplementary Figure 3: Diagnostic test accuracy of AI models in MRI in ROC space. The size of each marker is proportional to the size of the test dataset. An ideal classifier would be at the top-left corner at (0, 1).

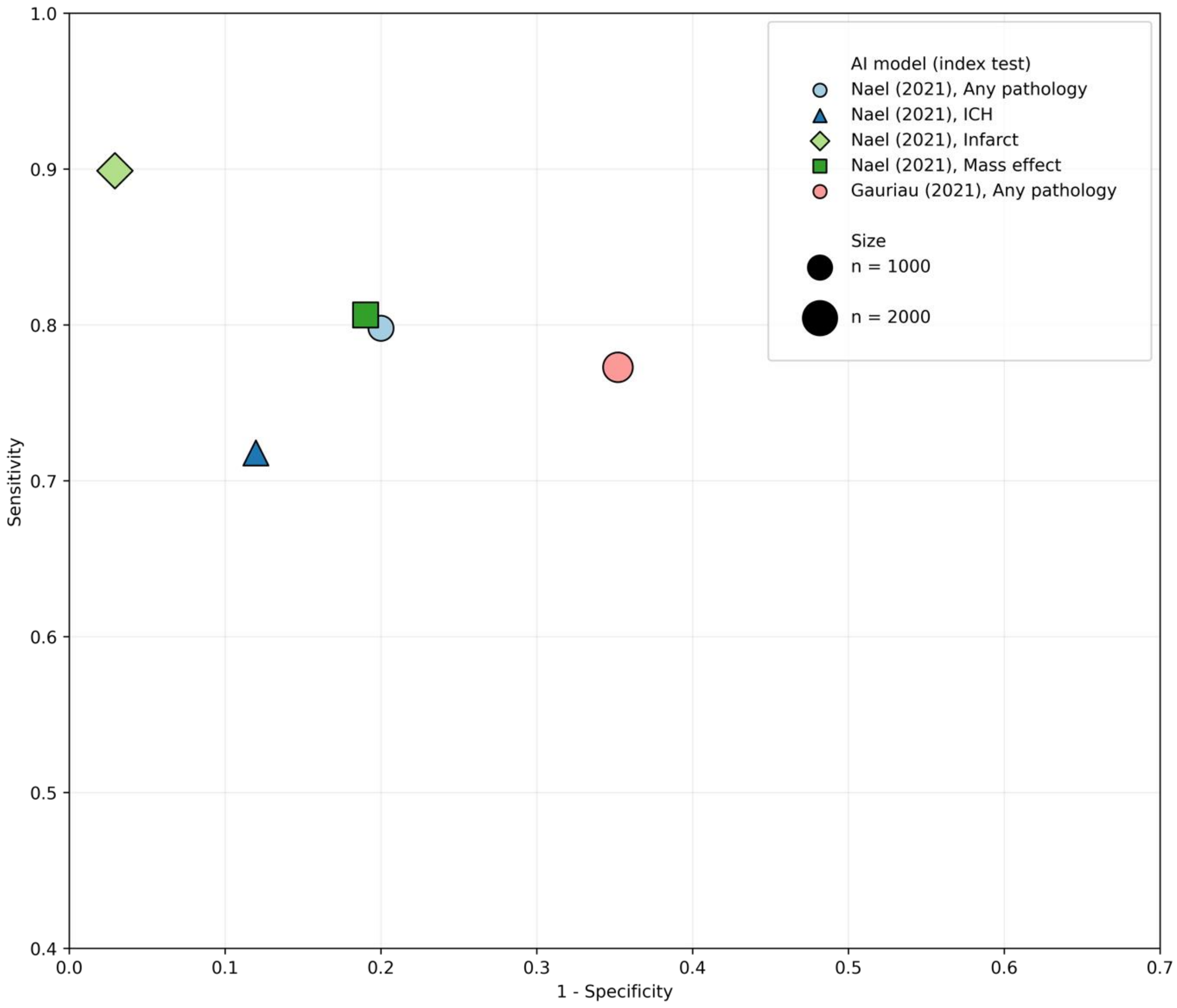